\documentclass{elsart}
\usepackage{epsf}
\def\sst{\scriptscriptstyle}

\begin{document}
\begin{frontmatter}
\title{Unitary-model-operator approach \\to ${\it \Lambda }$ hypernuclei}
\author[Kyushu]{Shinichiro Fujii}
\author[KIT]{Ryoji Okamoto}
\author[KIT]{Kenji Suzuki}

\address[Kyushu]{Department of Physics, Kyushu University,
Fukuoka 812-8581, Japan}
\address[KIT]{Department of Physics, Kyushu Institute of Technology,
Kitakyushu 804-8550, Japan}

\begin{abstract}

A method is formulated for the description of lambda hypernuclei
in the framework of the unitary-model-operator approach (UMOA).
The method is applied to
 $_{\it \Lambda }^{17}$O. A lambda-nucleon effective interaction
 is derived, taking the coupling of the sigma-nucleon channel into account.
The lambda single-particle energies are
 calculated for the
 0$s_{1/2}$, 0$p_{3/2}$ and 0$p_{1/2}$ states employing
the Nijmegen soft-core (NSC), J\"ulich model-\~A (J\~A)
and model-\~B (J\~B) hyperon-nucleon potentials.
\end{abstract}
\end{frontmatter}

\section{Introduction}
\setcounter{equation}{0}

There have been a number of attempts to study hypernuclei containing not only 
nucleons but also hyperons with strangeness
\cite{rf:GAL77,rf:POV78,rf:BAN85,rf:BAN90,rf:MAK94,rf:GIB95}.
It has been of high interest to investigate how the presence
 of strangeness broadens and/or modifies the knowledge achieved in the field of
 ordinary nuclear physics.

It is a fundamental problem to understand properties of a quantum many-body
system in terms of basic interactions between constituent particles.
In hypernuclear physics a unified treatment of all the
baryon-baryon interactions, especially nucleon-nucleon (${\it NN}$) and
hyperon-nucleon ($ {\it YN} $) interactions, will be necessary if one wish to
understand general properties of hypernuclei.
However, our knowledge 
on the $ {\it YN} $ interaction is quite inadequate in contrast to that 
of the $ {\it NN} $ interaction due to experimental difficulties.
In such a situation many-body theoretical studies of hypernuclei
could provide alternative information on the $ {\it YN} $ interaction. 

It has been discussed that 
physics of hypernuclei is different from conventional nuclear
 physics in some aspects. First, the channel couplings such as the 
$ {\it \Sigma N} $-$ {\it \Lambda N} $ coupling play 
a significant role in the structure of hypernuclei in contrast to
the very limited role of the ${\it \Delta N}$-${\it NN }$ coupling appeared in
nuclear many-body systems \cite{rf:GIB94}.
Second, the anti-symmetric spin-orbit force, which is forbidden in 
ordinary nuclei, emerges in hypernuclei in addition to the symmetric spin-orbit force.

The spin-orbit splitting of $ {\it \Lambda} $ single-particle levels in 
$ {\it \Lambda} $ hypernuclei
was considered experimentally to be much smaller
than their nucleonic counterpart \cite{rf:BRU76}.
However, some recent studies as in Ref.~\cite{rf:DDMT}
suggest that the splitting would not be so small as discussed
 earlier. It will be highly desirable for the spin-orbit splitting to be 
established in experiment. The magnitude of the spin-orbit splitting will put 
further constraints on the ${\it YN}$ interaction.

There have been several possibilities of
solving the structure of hypernuclei.
One possibility is to 
treat few-body or light hypernuclei by solving the Faddeev
equation \cite{rf:MKGS}
or by applying the cluster model \cite{rf:YOMI94,rf:HKMYY}.
Another possibility is microscopic studies of hypernuclei with a larger mass 
number.
While a few lower partial waves play a 
decisive role in few-body or light hypernuclei, higher partial waves could 
give some contributions in heavier hypernuclei. These microscopic studies of 
hypernuclei could make us understand interesting physics
of their structure itself, and also help us in determining the 
$ {\it YN} $ interaction.  

Microscopic structure calculation of heavier 
hypernuclei is usually performed by introducing a ${\it YN}$ effective
 interaction.
For this purpose the $G$-matrix theory has been applied in two ways:
In some works
an effective $ {\it YN} $ interaction has been derived by simulating a
nuclear matter $G$-matrix.
In this approach the non-locality and energy dependence of the $G$-matrix are
neglected, and the effective $ {\it YN} $ interaction is represented in 
the two-
or three-range Gaussian form with the adjustable Fermi momentum $k_{\rm F}$.
This type of the effective interaction has been referred to as the YNG
interaction \cite{rf:YB85}. The YNG interaction has been successfully
used in many of hypernuclear structure calculations \cite{rf:YMHIN}.
It seems to us that the YNG interaction
has made an important contribution to the considerable progress in overall
description of hypernuclei.
The other approach, in the framework of the $G$-matrix theory, is to calculate
the $G$-matrix in finite hypernuclei by treating rigorously the Pauli exclusion
operator and making a self-consistent calculation of single-particle potential
of hyperons \cite{rf:BAN81,rf:HAL93,rf:HAO93,rf:HJO96,rf:VID98}.
In contrast to studies of using the YNG interaction
this approach possibly gives a better description of the state dependence
of the effective $ {\it YN} $ interaction.
It has been established that the $G$-matrix can be used as a basic ingredient
in constructing a microscopic many-body theory starting with bare interactions.

In spite of a marked success in the applications of the $G$-matrix theory,
we may say from a formal point of view that the $G$-matrix is
energy-dependent and
non-hermitian, and does not have the property of decoupling between 
a model space of low-lying two-particle states and its complement.
The $G$-matrix is not considered to be an effective 
interaction in a formal sense of the effective interaction theory.
In order to derive an effective
interaction we should add some higher-order corrections
such as folded diagrams \cite{rf:KO,rf:HAO96}.
It would be desirable if we could have a theory describing
many-body systems in terms of an energy-independent and hermitian 
effective interaction with the property of decoupling. Two of the present 
authors, K.~S. and R.~O., proposed a many-body theory, 
the unitary-model-operator approach (UMOA)
that was formulated on the basis of such an effective interaction.
The UMOA was applied to finite nuclei,
$^{16}$O and $^{40}$Ca \cite{rf:SO,rf:KSO}, and 
some improvements have been attained.
The method may be viewed as an alternative way of deriving an 
effective interaction in finite many-body systems.

A unified study of hypernuclei and ordinary nuclei
 would be helpful in understanding many-body systems of baryons
 in a microscopic way. Therefore, we extended 
the formulation of the UMOA, in the previous work \cite{rf:FOS},
to the description of hypernuclei and made a calculation of
the properties in
$^{17}_{\it \Lambda }$O with a realistic ${\it YN}$ interaction.
The purposes of this work are to present a general formulation of the
extended UMOA and apply it to $^{17}_{\it \Lambda }$O
by employing some realistic ${\it YN}$ interactions.

This paper is organized as follows: In Section 2 we present a general
formulation of the UMOA for deriving  the ${\it \Lambda N}$
effective interaction. In Section 3 the approximation procedure is given
for making actual calculations.
In Section 4 we apply the UMOA to $^{17}_{\it \Lambda }$O
by employing the Nijmegen
soft-core (NSC) \cite{rf:NSC}, J\"ulich model-\~A (J\~A)
and model-\~B (J\~B) \cite{rf:JUL}
${\it YN}$ potentials.
In Section 5 we make some concluding remarks.

\section{Formulation of the UMOA for the $ {\it \Lambda N} $ effective
interaction}
\setcounter{equation}{0}

We present a formulation of the UMOA for applying to a calculation of the
properties in $ {\it \Lambda} $ hypernuclei, especially, closed-shell
nucleus plus one $ {\it \Lambda} $ systems.
A main purpose is to give a method of calculating an effective interaction
between $ {\it \Lambda} $ and $ {\it N} $.
In the derivation of the $ {\it \Lambda N} $ effective interaction we should
note that, because of the small mass difference of about 77MeV between
$ {\it \Lambda} $ and $ {\it \Sigma} $, the coupling of the
$ {\it \Lambda N} $  and $ {\it \Sigma N} $ channels plays an important role.

We first consider a two-body hamiltonian of a
$ {\it \Sigma N} $-$ {\it \Lambda N} $ coupled system given by
\begin{eqnarray}
\label{eq:LHYN}
H_{\sst {\it YN}}=h_{\sst {\it YN}}+\Delta m+v_{\sst {\it YN}},
\end{eqnarray}
where
\begin{eqnarray}
\label{eq:SHYN}
  h_{\sst {\it YN}}=
  \left( \begin{array}{cc}
             t_{\sst {\it \Lambda }}+u_{\sst {\it \Lambda }}+t_{\sst {\it N}}
+u_{\sst {\it N}} &       0  \\
                                  0         &    t_{\sst {\it \Sigma }}
+u_{\sst {\it \Sigma }}+t_{\sst {\it N}}+u_{\sst {\it N}}
           \end{array} \right),
\end{eqnarray}
\begin{eqnarray}
\label{eq:DM}
  \Delta m=\left( \begin{array}{cc}
               0         &   0 \\
               0         &   m_{\sst {\it \Sigma }}-m_{\sst {\it \Lambda }}
           \end{array} \right)
\end{eqnarray}
and
\begin{eqnarray}
\label{eq:VYN}
  v_{\sst {\it YN}}=\left( \begin{array}{cc}
             v_{\sst {\it \Lambda N-\Lambda N}}  &  
v_{\sst {\it \Lambda N-\Sigma N}}  \\
             v_{\sst {\it \Sigma N-\Lambda N}}    &  
v_{\sst {\it \Sigma N-\Sigma N}}
           \end{array} \right).
\end{eqnarray}
The terms $ t_{k} $ and  $ u_{k} $ for $ k= {\it \Lambda , \Sigma} $ and
$ {\it N} $
are the kinetic and single-particle potential energies of
$ {\it \Lambda, \Sigma} $
and $ {\it N} $, respectively, and $ v_{\sst {\it YN}} $ is
the bare two-body ${\it YN}$ interaction
including the $ {\it \Lambda N} $ and $ {\it \Sigma N} $ channels.
The terms $ m_{\sst {\it \Lambda }} $ and
$ m_{\sst {\it \Sigma }} $ are the rest masses of
$ {\it \Lambda } $ and $ {\it \Sigma } $, respectively.
The hamiltonian $H_{\sst {\it YN}}$ describes a subsystem of
interacting two particles $ {\it \Lambda N} $ or $ {\it \Sigma N} $ in the
one-body potentials $u_{\sst {\it \Lambda }}$,
$u_{\sst {\it \Sigma }}$ and $u_{\sst {\it N}}$.

In order to derive a $ {\it \Lambda N} $ effective interaction
we introduce a unitarily transformed $ {\it YN} $ interaction given by 
\begin{eqnarray}
\label{eq:TVYN}
  \tilde{v}_{\sst {\it YN}}=e^{-S_{\it YN}}(h_{\sst {\it YN}}+\Delta m
+v_{\sst {\it YN}})e^{S_{\it YN}}-(h_{\sst {\it YN}}+\Delta m),
\end{eqnarray}
where $ S_{\sst {\it YN}} $ is the correlation operator defined
in a space of two-particle states of $ {\it \Lambda N} $ and
$ {\it \Sigma N} $.
The $ S_{\sst {\it YN}} $ is an anti-hermitian operator
satisfying
\begin{eqnarray}
S_{\sst {\it YN}}^{\dagger }=-S_{\sst {\it YN}}.
\end{eqnarray}
In the formulation of the UMOA the correlation operator
$ S_{\sst {\it YN}} $ is determined by the equation of decoupling for
$\tilde{v}_{\sst {\it YN}}$
between a certain model space consisting of low-momentum two-particle states
and its complement.
This determination of the correlation operator
$ S_{\sst {\it YN}} $
is one of the characteristics of the present approach.

For the determination of a $ {\it \Lambda N} $ effective interaction,
we introduce a model space consisting of low-momentum $ {\it \Lambda N} $
states.
Let $P_{\sst {\it \Lambda N}}$ be the projection operator onto the
model space.
The complement of the $P_{\sst {\it \Lambda N}} $ space includes a
space of high-momentum $ {\it \Lambda N} $ states referred to as the
$Q_{\sst {\it \Lambda N}}$
space and a space of all the $ {\it \Sigma N} $ states referred to as the
$Q_{\sst {\it \Sigma N}}$ space.
The condition of decoupling for the transformed interaction
$\tilde{v}_{\sst {\it YN}}$ is given by
\begin{eqnarray}
\label{eq:DEC}
(Q_{\sst {\it \Lambda N}}+Q_{\sst {\it \Sigma N}})\tilde{v}_{\sst {\it YN}}
P_{\sst {\it \Lambda N}}=P_{\sst {\it \Lambda N}}\tilde{v}_{\sst {\it YN}}
(Q_{\sst {\it \Lambda N}}+Q_{\sst {\it \Sigma N}})=0.
\end{eqnarray}
Hereafter, we refer to the above equation as the decoupling equation.
We assume that the one-body part $ h_{\sst {\it YN}}+\Delta m $
in Eq.~(\ref{eq:TVYN}) is decoupled as
\begin{eqnarray}
\label{eq:ASSUME}
(Q_{\sst {\it \Lambda N}}+Q_{\sst {\it \Sigma N}})
(h_{\sst {\it YN}}+\Delta m)P_{\sst {\it \Lambda N}}
=P_{\sst {\it \Lambda N}}(h_{\sst {\it YN}}+\Delta m)
(Q_{\sst {\it \Lambda N}}+Q_{\sst {\it \Sigma N}})=0.
\end{eqnarray}
On the above assumption the decoupling equation (\ref{eq:DEC}) becomes
\begin{eqnarray}
\label{eq:DECR}
(Q_{\sst {\it \Lambda N}}+Q_{\sst {\it \Sigma N}})
e^{-S_{\it YN}}(h_{\sst {\it YN}}
+\Delta m+v_{\sst {\it YN}})e^{S_{\it YN}}P_{\sst {\it \Lambda N}}=0.
\end{eqnarray}
In general, the decoupling equation (\ref{eq:DECR}) does not
determine $ S_{\sst {\it YN}} $ uniquely. 
The usual restrictive conditions are 
\begin{eqnarray}
\label{eq:REST}
P_{\sst {\it \Lambda N}}S_{\sst {\it YN}}P_{\sst {\it \Lambda N}}
=(Q_{\sst {\it \Lambda N}}+Q_{\sst {\it \Sigma N}})S_{\sst {\it YN}}
(Q_{\sst {\it \Lambda N}}+Q_{\sst {\it \Sigma N}})=0.
\end{eqnarray}
These conditions are called the minimal effect requirements \cite{rf:Bran},
which mean that
the transformation exp$(S_{\sst {\it YN}})$ does not induce
unnecessary transformation within each of
the $ P_{\sst {\it \Lambda N}} $ and $ Q_{\sst {\it \Lambda N}}
+Q_{\sst {\it \Sigma N}} $ spaces.

In the general theory of constructing a hermitian effective interaction,
the solution for $S_{\sst {\it YN}}$ in Eq.~(\ref{eq:DECR}) has been
known \cite{rf:SR80,rf:West81,rf:Su82} and is given by
\begin{eqnarray}
\label{eq:SYN}
S_{\sst {\it YN}}
&=& \arctan \hspace{-0.9mm} {\rm h} (\omega _{\sst {\it YN}}
-\omega _{\sst {\it YN}}^{\scriptstyle \dag }) \nonumber \\
&=& \sum_{n=0}^{\infty }\frac{(-1)^{n}}{2n+1}\{ \omega _{\sst {\it YN}}
(\omega _{\sst {\it YN}}^{\dag } \omega _{\sst {\it YN}})^{n}-\rm h.c. \}
\end{eqnarray}
with
\begin{eqnarray}
\label{eq:OYN}
\omega _{\sst {\it YN}}=\sum_{k=1}^{d}(Q_{\sst {\it \Lambda N}}
+Q_{\sst {\it \Sigma N}})
|\Psi _{{\sst {\it YN}}k}\rangle
\langle \overline{\psi_{{\sst {\it \Lambda N}}k}}
|P_{\sst {\it \Lambda N}},
\end{eqnarray}
where $ d $ is the dimension of the $ P_{\sst {\it \Lambda N}} $ space.
The $ |\Psi _{{\sst {\it YN}}k}\rangle $
is the two-body eigenstate consisting of two components of $ {\it \Lambda N} $
and $ {\it \Sigma N} $ states, and satisfies

\begin{eqnarray}
\label{eq:EIGE}
H_{\sst {\it YN}}
|\Psi _{{\sst {\it YN}}k}\rangle =E_{k}|\Psi _{{\sst {\it YN}}k}\rangle , 
\end{eqnarray}
where
\begin{eqnarray}
\label{eq:EIGS}
|\Psi _{{\sst {\it YN}}k}\rangle=
\left[ \begin{array}{c}
           |\Psi _{{\sst {\it \Lambda N}}k}\rangle \\
           |\Psi _{{\sst {\it \Sigma N}}k}\rangle
         \end{array} \right].
\end{eqnarray}
The $ \langle \overline{\psi_{{\sst {\it \Lambda N}}k} }| $ in
Eq.~(\ref{eq:OYN}) is the biorthogonal state of
$ |\psi_{{\sst {\it \Lambda N}}k} \rangle $ satisfying
\begin{eqnarray}
\label{eq:BIO}
\langle \overline{\psi_{{\sst {\it \Lambda N}}k} }
|\psi_{{\sst {\it \Lambda N}}k'}\rangle =\delta_{kk'},
\end{eqnarray} 
where $ |\psi_{{\sst {\it \Lambda N}}k} \rangle $ is the
$ P_{\sst {\it \Lambda N}} $-space component of
$ |\Psi _{{\sst {\it YN}}k}\rangle $ defined by
\begin{eqnarray}
\label{eq:PSC}
|\psi_{{\sst {\it \Lambda N}}k} \rangle =P_{\sst {\it \Lambda N}}
|\Psi _{{\sst {\it YN}}k}\rangle .
\end{eqnarray}
It should be noted that the solution $ S_{\sst {\it YN}} $ in
Eq.~(\ref{eq:SYN}) is determined dependently
on the choice of a set of $ d $ eigenstates
$ \{ |\Psi _{{\sst {\it YN}}k}\rangle , k=1, 2, ... ,d \} $ as given
in Eq.~(\ref{eq:OYN}). 
We choose a set of $ \{ |\Psi _{{\sst {\it YN}}k}\rangle \} $
so that they have the largest $ P_{\sst {\it \Lambda N}} $-space overlaps
among all the eigenstates in Eq.~(\ref{eq:EIGE}).
This is a usual choice in the derivation of the effective interaction.

With the solution $ S_{\sst {\it YN}} $ in Eq.~(\ref{eq:SYN}),
the $ {\it \Lambda N} $ effective interaction in the
$ P_{\sst {\it \Lambda N}} $ space is given by 
\begin{eqnarray}
\label{eq:TVYNP}
\tilde{v}_{\sst {\it \Lambda N}}=P_{\sst {\it \Lambda N}}
\tilde{v}_{\sst {\it YN}}P_{\sst {\it \Lambda N}}.
\end{eqnarray}
It has been known that this $ {\it \Lambda N} $ effective interaction
can be written explicitly \cite{rf:SO,rf:KEHLSOK} as
\begin{eqnarray}
\label{eq:EVLN}
\hspace{-5mm}\langle \phi _{{\sst {\it \Lambda N}}\alpha }
|\tilde{v}_{\sst {\it \Lambda N}}
|\phi_{{\sst {\it \Lambda N}}\beta }\rangle =
\frac{(1+\mu _{\alpha }^{2})^{\frac{1}{2}}
\langle \phi _{{\sst {\it \Lambda N}}\alpha }
|R_{\sst {\it \Lambda N}}|\phi_{{\sst {\it \Lambda N}}\beta }\rangle 
+(1+\mu _{\beta }^{2})^{\frac{1}{2}}
\langle \phi _{{\sst {\it \Lambda N}}\alpha }
|R_{\sst {\it \Lambda N}}^{\dag }
|\phi_{{\sst {\it \Lambda N}}\beta }\rangle }
{(1+\mu _{\alpha }^{2})^{\frac{1}{2}}+(1+\mu _{\beta }^{2})^{\frac{1}{2}}},
\nonumber \\
\end{eqnarray}
where $ |\phi_{{\sst {\it \Lambda N}}\alpha }\rangle $
($ |\phi_{{\sst {\it \Lambda N}}\beta }\rangle $)
and $ \mu _{\alpha} $ ($ \mu _{\beta }$) are defined
through the eigenvalue equation
\begin{eqnarray}
\label{eq:OEIGE}
\omega _{\sst {\it YN}}^{\dag } \omega _{\sst {\it YN}}
|\phi_{{\sst {\it \Lambda N}}\alpha }\rangle
=\mu _{\alpha }^{2}|\phi_{{\sst {\it \Lambda N}}\alpha }\rangle ,
\end{eqnarray}
and $ R_{\sst {\it \Lambda N}} $ is given by
\begin{eqnarray}
\label{eq:NHEVLN}
R_{\sst {\it \Lambda N}}=P_{\sst {\it \Lambda N}}
(v_{\sst {\it YN}}+v_{\sst {\it YN}}\omega _{\sst {\it YN}})
P_{\sst {\it \Lambda N}}.
\end{eqnarray}
The $R_{\sst {\it \Lambda N}}$ is a
$P_{\sst {\it \Lambda N}}$-space operator that agrees with the
effective interaction of non-hermitian type.
Equation (\ref{eq:EVLN}) is a formula for converting
the non-hermitian effective interaction to the hermitian one.

The term $ u_{\sst {\it \Lambda }} $ in Eq.~(\ref{eq:SHYN}) is
introduced as a self-consistent potential of $ {\it \Lambda } $ calculated with
 the $ {\it \Lambda N} $ effective interaction
 $ \tilde{v}_{\sst {\it \Lambda N}} $ as 
\begin{eqnarray}
\label{eq:UL}
\langle \alpha _{\sst {\it \Lambda }}|u_{\sst {\it \Lambda }}
|\alpha _{\sst {\it \Lambda }}' \rangle =\sum_{\xi _{\sst {\it N}}
:{\rm occupied}}
\langle \alpha _{\sst {\it \Lambda }}\xi _{\sst {\it N}}
|\tilde{v}_{\sst {\it \Lambda N}}
|\alpha _{\sst {\it \Lambda }}'\xi _{\sst {\it N}}\rangle ,
\end{eqnarray}
where $ |\alpha _{\sst {\it \Lambda }}\rangle $
($ |\alpha _{\sst {\it \Lambda }}'\rangle $)
and $ |\xi _{\sst {\it N}}\rangle $ are the single-particle states of
 $ {\it \Lambda } $ and $ {\it N} $, respectively. 

From Eqs.~(\ref{eq:EVLN})-(\ref{eq:NHEVLN}) we see that
the $ {\it \Lambda N} $ effective
interaction $\tilde{v}_{\sst {\it \Lambda N}}$ can be calculated
once the operator $ \omega _{\sst {\it YN}} $ is given.
The operator $ \omega _{\sst {\it YN}} $ is
determined completely,
as given in Eq.~(\ref{eq:OYN}), by solving the eigenvalue
equation (\ref{eq:EIGE}) and selecting $ d $ eigenstates
$ \{ |\Psi _{{\sst {\it YN}}k}\rangle \} $ with the
largest $ P_{\sst {\it \Lambda N}} $-space overlaps.
It is noted here that Eqs.~(\ref{eq:SYN})-(\ref{eq:UL}) give a set of
equations that determines the $ {\it \Lambda N} $ effective interaction
$ \tilde{v}_{\sst {\it \Lambda N}} $ self-consistently with the
single-particle potential $ u_{\sst {\it \Lambda }} $.

\section{Approximation procedure}
\subsection{Two-step method for  the calculation of
 $\tilde{v}_{\sst {\it YN}}$}
\setcounter{equation}{0}

Here we discuss a calculation procedure for obtaining the $ {\it \Lambda N} $
effective interaction acting in a small model space of low-lying
$ {\it \Lambda N} $ states.
We use the harmonic oscillator (h.o.) wave functions as basis states.
It is desirable that the single-particle potentials of ${\it \Lambda}$,
$ {\it \Sigma } $ and $ {\it N} $ in intermediate
states are taken into account in a self-consistent way. However, as our
main concern is to derive a $ {\it \Lambda N} $  effective interaction,
the self-consistent treatment of $u_{\sst {\it \Lambda }}$ would have
predominant importance in the present calculation.
For this reason we make a
self-consistent calculation only for $u_{\sst {\it \Lambda }}$,
and as for $u_{\sst {\it N}}$ we employ
a fixed potential calculated previously for closed shell nuclei
in the UMOA \cite{rf:SO}. The effect of the single-particle potential
$u_{\sst {\it \Sigma }}$ on the $ {\it \Lambda N} $
effective interaction has not necessarily been clarified,
although some calculations with $u_{\sst {\it \Sigma }}$ have
been made for the $G$-matrices in hyperonic nuclear matter
\cite{rf:YB90,rf:SLCBL95}.
In the actual calculations made so far the single-particle potential
$u_{\sst {\it \Sigma }}$
has been considered to be rather shallow \cite{rf:YMHIN}.
Although we have not known clearly how much the effect of
$u_{\sst {\it \Sigma }}$ on the $ {\it \Lambda N} $
effective interaction, we here assume $u_{\sst {\it \Sigma }}=0$.

In solving the self-consistent equation for $u_{\sst {\it \Lambda }}$
with $\tilde{v}_{\sst {\it \Lambda N}}$
we should be careful in the following two respects:
(a) In the calculation of energies of intermediate states in propagators
we should use $u_{\sst {\it \Lambda }}$ that is given for 
${\it \Lambda}$ states including high-momentum states.  In order to do this we
need a $ {\it \Lambda N} $ effective interaction acting in a large model space,
because in general the 
$Q_{\sst {\it \Lambda N}}$ space contains high-momentum
$ {\it \Lambda N} $
states and the $Q$-space effective interaction
$ Q_{\sst {\it \Lambda N}}\tilde{v}_{\sst {\it YN}}
Q_{\sst {\it \Lambda N}}$ is not always well behaved. We thus make a
calculation by introducing two model spaces, namely, large and small model
spaces.  The $ {\it \Lambda N} $ effective interaction in each of the model
spaces is calculated by following a  two-step procedure.
In the first-step procedure,
we define a large model space and calculate the $ {\it \Lambda N} $ effective
interaction and single-particle potential of ${\it \Lambda}$
in high-lying states.
The $ {\it \Lambda N} $ effective interaction acting in this large model
space will not be suitable for the effective interaction in a usual sense.
By introducing a small model space and
using the $ {\it \Lambda N} $ effective interaction determined
in the first-step
calculation, we proceed to the second-step calculation where we calculate a
$ {\it \Lambda N} $ effective interaction acting in this small model space of
low-lying $ {\it \Lambda N}$ states.  
(b) It is important in actual calculations how to determine the $P$ and $Q$
spaces
because an effective interaction is determined dependently on the choice of
the $P$ and $Q$ spaces. If a state in the $P$ space mixes strongly with
$Q$-space
states in the eigenstate in Eq.~(\ref{eq:EIGE}), the matrix element of
$\omega_{\sst {\it YN}}$ or $S_{\sst {\it YN}}$ defined in
Sec.~2 becomes large and as a result higher-order many-body
correlations \cite{rf:SO86} give rise to non-negligible contributions.
In general, the
strong mixing takes place when some of the $P$- and $Q$-space states are
quasi-degenerate in energy. For this reason, we should choose
the $P$ and $Q$ spaces which are well separated in energy.

With due regard to points (a) and (b) we perform a calculation of the
effective interaction through the two-step procedure.
In the following subsections we present the two-step method in detail.

\subsection{First-step decoupling}

Let us write a two-particle state of ${\it \Lambda}$ and $ {\it N} $ as the
product of the h.o. 
single-particle wave functions in the usual notation as
\begin{eqnarray}
\label{eq:TPS}
|\alpha _{\sst {\it Y_{i}}}\beta _{\sst {\it N}}\rangle 
=|n_{\sst {\it Y_{i}}}
  l_{\sst {\it Y_{i}}}
  j_{\sst {\it Y_{i}}}
  m_{\sst {\it Y_{i}}},
  n_{\sst {\it N}}
  l_{\sst {\it N}}
  j_{\sst {\it N}}
  m_{\sst {\it N}}\rangle,\ ({\it Y_{i}}={\it \Lambda, \Sigma}).
\end{eqnarray}
We define a model space denoted by 
$P_{\sst {\it \Lambda N}}^{\sst (1)}$ 
and its complement
$ Q_{\sst {\it \Lambda N}}^{\sst (1)} $ 
with a boundary number $ \rho _{\sst 1} $ as
\begin{eqnarray}
\label{eq:DTPS}
|\alpha _{\sst {\it \Lambda }}\beta _{\sst {\it N}}\rangle 
\in \left\{ 
\begin{array}{ll}
\displaystyle P_{\sst {\it \Lambda N}}^{\sst (1)} &
{\rm if \ } 2n_{\sst {\it \Lambda }}+l_{\sst {\it \Lambda }}
+2n_{\sst {\it N}}+l_{\sst {\it N}} \leq \rho _{\sst 1}, \\
\displaystyle Q_{\sst {\it \Lambda N}}^{\sst (1)} & {\rm otherwise}.\\
\end{array}\right.
\end{eqnarray}
We also define a space of all the $ {\it \Sigma N} $ states denoted by 
$Q_{\sst {\it \Sigma N}}^{\sst (1)}$ as
\begin{eqnarray}
\label{eq:STPS}
|\alpha _{\sst {\it \Sigma }}\beta _{\sst {\it N}}\rangle 
\in Q_{\sst {\it \Sigma N}}^{\sst (1)}.
\end{eqnarray}
Note that the model space 
$P_{\sst {\it \Lambda N}}^{\sst (1)}$
does not contain the $ {\it \Sigma N} $ states.

If we make a graph of two-particle states, the 
$ P_{\sst {\it \Lambda N}}^{\sst (1)} $, 
$ Q_{\sst {\it \Lambda N}}^{\sst (1)} $ and 
$ Q_{\sst {\it \Sigma N}}^{\sst (1)} $ spaces
are represented in Fig.~1.
The $ \rho _{\sst 1} $ means a number such that
$ (\rho _{\sst 1}+3) {\it \hbar \Omega }/2$ gives the maximum
kinetic energy of the $ {\it \Lambda N} $ two-particle states in the 
$ P_{\sst {\it \Lambda N}}^{\sst (1)} $ space.
The $ {\it \hbar \Omega } $ denotes the unit of the h.o. energy
which we use commonly for $ {\it \Lambda } $, $ {\it \Sigma } $
and $ {\it N} $.
As we require to introduce a sufficiently large model space 
in the first-step calculation,
we should choose $ \rho _{\sst 1} $ as large as possible.
Another requirement is that the states in the
$P_{\sst {\it \Lambda N}}^{\sst (1)}$ space should be
separated in energy from the states in the
$Q_{\sst {\it \Lambda N}}^{\sst (1)}$ space.
For this reason, we adopt a model space
$ P_{\sst {\it \Lambda N}}^{\sst (1)} $
of a triangular shape. This choice of the
$ P_{\sst {\it \Lambda N}}^{\sst (1)} $ space
guarantees the separation in energy.

\begin{figure*}
  \label{fig:1}
  \epsfxsize = 13cm
  \centerline{\epsfbox{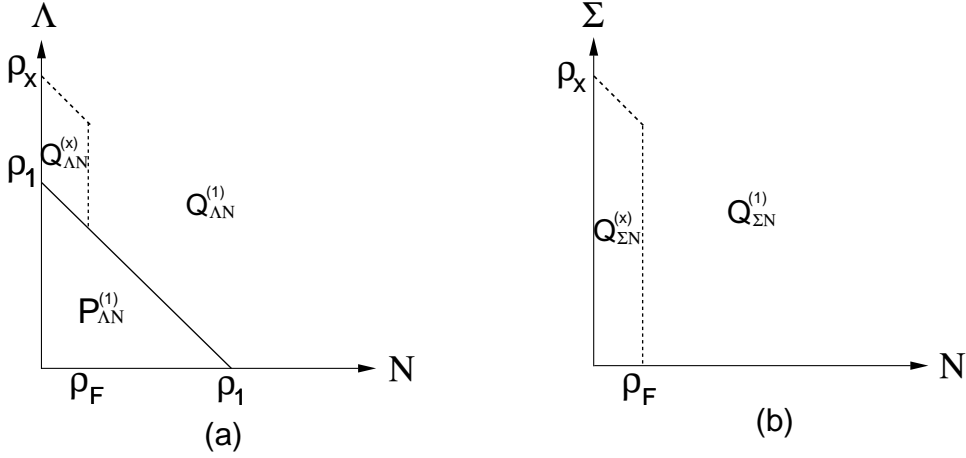}}
  \caption {The model space and its complement in the
 first-step calculation.}
\end{figure*}

In Fig.~1 the spaces 
$Q^{\rm \sst (X)}_{\sst {\it Y_{i}N}}$
for $ {\it Y_{i}} = {\it \Lambda} $ and $ {\it \Sigma } $, specified by
the numbers $\rho _{\sst 1},\rho _{\rm \sst F}$ and 
$\rho _{\rm \sst X}$, stand for the excluded spaces due to the
Pauli principle for nucleons.
The number $\rho _{\rm \sst F}$ is defined as 
$\rho _{\rm \sst F}=
2n_{\sst {\it N}}+l_{\sst {\it N}}$
with the quantum numbers
$\{n_{\sst {\it N}},l_{\sst {\it N}}\}$
of the highest occupied orbit in the closed-shell core nucleus.
The $ \rho _{\rm \sst X} $ means a number such that
$ (\rho _{\rm \sst X}+3) {\it \hbar \Omega }/2$ gives the maximum
kinetic energy of the ${\it YN}$ two-particle states in the 
$Q_{\sst {\it \Lambda N}}^{\rm \sst (X)} $ and
$Q_{\sst {\it \Sigma N}}^{\rm \sst (X)} $ 
spaces. We then write the 
$Q_{\sst {\it \Lambda N}}^{\rm \sst (X)} $ and
$Q_{\sst {\it \Sigma N}}^{\rm \sst (X)} $ 
spaces as
\begin{eqnarray}
\label{eq:TPSQL}
|\alpha _{\sst {\it \Lambda }}\beta _{\sst {\it N}}\rangle 
\in Q_{\sst {\it \Lambda N}}^{\rm \sst (X)}\ \ 
&&\ {\rm if \ } \rho _{\sst 1} < 
 2n_{\sst {\it \Lambda }}+l_{\sst {\it \Lambda }}
+2n_{\sst {\it N}}+l_{\sst {\it N}} 
 \leq \rho _{\rm \sst X} \nonumber \\
&&\hspace{-1mm}{\rm and \ } 0 \leq 2n_{\sst {\it N}}
+l_{\sst {\it N}} 
\leq \rho _{\rm \sst F}
\end{eqnarray}
and
\begin{eqnarray}
\label{eq:TPSQS}
|\alpha _{\sst {\it \Sigma }}\beta _{\sst {\it N}}\rangle 
\in Q_{\sst {\it \Sigma N}}^{\rm \sst (X)}\ \ 
&&\ {\rm if \ } 0 \leq
 2n_{\sst {\it \Sigma }}+l_{\sst {\it \Sigma }}
+2n_{\sst {\it N}}+l_{\sst {\it N}} 
 \leq \rho _{\rm \sst X} \nonumber \\
&&\hspace{-2.2mm}{\rm and \ } 0 \leq 2n_{\sst {\it N}}
+l_{\sst {\it N}} 
\leq \rho _{\rm \sst F}.
\end{eqnarray}
We also define the sum of the excluded spaces as
\begin{eqnarray}
\label{eq:QYNX}
Q^{\rm \sst (X)}_{\sst {\it YN}} 
 \equiv
   Q^{\rm \sst (X)}_{\sst {\it \Lambda N}}
 + Q^{\rm \sst (X)}_{\sst {\it \Sigma N}}.
\end{eqnarray}
Furthermore, by using the projection operators,
$Q_{\sst {\it \Lambda N}}^{\sst (1)}$,
$Q_{\sst {\it \Lambda N}}^{\rm \sst (X)}$,
$Q_{\sst {\it \Sigma N}}^{\sst (1)}$ and
$Q_{\sst {\it \Sigma N}}^{\rm \sst (X)}$,
we introduce projection operators onto the spaces of allowed ${\it YN}$
states as
\begin{eqnarray}
\label{eq:QLNB}
\overline{Q}_{\sst {\it \Lambda N}}^{\sst (1)} 
 \equiv Q^{\sst (1)}_{\sst {\it \Lambda N}}
 - Q^{\rm \sst (X)}_{\sst {\it \Lambda N}},
\end{eqnarray}
\begin{eqnarray}
\label{eq:QSNB}
 \overline{Q}_{\sst {\it \Sigma N}}^{\sst (1)} 
 \equiv Q_{\sst {\it \Sigma N}}^{\sst (1)}
 - Q^{\rm \sst (X)}_{\sst {\it \Sigma N}},
\end{eqnarray}
and we also define
\begin{eqnarray}
\label{eq:QYNB}
\overline{Q}_{\sst {\it YN}}^{\sst (1)} 
 \equiv 
   \overline{Q}^{\sst (1)}_{\sst {\it \Lambda N}}
 + \overline{Q}^{\sst (1)}_{\sst {\it \Sigma N}}.
\end{eqnarray}

Considering the Pauli-exclusion principle for nucleons the terms
$Q_{\sst {\it YN}}^{\rm \sst (X)}
 v_{\sst {\it YN}}
 Q_{\sst {\it YN}}^{\rm \sst (X)}$ and
$P_{\sst {\it \Lambda N}}^{\sst (1)}
 v_{\sst {\it YN}}
 Q_{\sst {\it YN}}^{\rm \sst (X)}+{\rm h.c.}$
should be removed in treating the problem of two-body correlations.
However, if we remove all the matrix elements of these terms, this leads to
overmuch counting of the Pauli principle. The matrix elements which are
diagonal in nucleon states should be restored even though nucleons are
in occupied states.
Therefore, instead of $v_{\sst {\it YN}}$, we should use a 
${\it YN}$ interaction $\overline{v}_{\sst {\it YN}}$ defined as 
\begin{eqnarray}
\label{eq:VBYN}
\overline{v}_{\sst {\it YN}}
 \equiv
  (P^{\sst (1)}_{\sst {\it \Lambda N}}
   +\overline{Q}^{\sst (1)}_{\sst {\it YN}})
  v_{\sst {\it YN}}
  (P^{\sst (1)}_{\sst {\it \Lambda N}}
   +\overline{Q}^{\sst (1)}_{\sst {\it YN}})
 + \sum_{\nu_{N}}
   \sum_{\stackrel{{\sst {\it Y_{i}}= {\it \Lambda, \Sigma }}}
        {{\sst {\it Y_{j}}={\it \Lambda, \Sigma }}}}
    Q^{\rm \sst (d)}_{\sst {\it Y_{i}N}}
    (\nu_{\sst {\it N}})
    v_{\sst {\it YN}}
    Q^{\rm \sst (d)}_{\sst {\it Y_{j}N}}
   (\nu_{\sst {\it N}}), \nonumber \\
\end{eqnarray}
where $\nu_{\sst {\it N}}$ denotes the set of the h.o. quantum numbers
 $\{ n_{\sst {\it N}},l_{\sst {\it N}}\} $
of an occupied state of a nucleon, and the operator 
$Q^{\rm \sst (d)}_{\sst {\it Y_{i}N}}
               (\nu_{\sst {\it N}})$ is defined as 
\begin{eqnarray}
\label{eq:QYND}
Q^{\rm \sst (d)}_{\sst {\it Y_{i}N}}
               (\nu_{\sst {\it N}})
\equiv
\sum_{\alpha_{\sst {\it Y_{i}}}}
       |\alpha_{\sst {\it Y_{i}}} \nu_{\sst {\it N}}\rangle
       \langle \alpha_{\sst {\it Y_{i}}} \nu_{\sst {\it N}}|,
      \ ({\it Y_{i}}= {\it \Lambda, \Sigma }).
\end{eqnarray}

The two-body ${\it YN}$ equation to be solved
in the first-step calculation is now written as
\begin{eqnarray}
\label{eq:EIGE1}
  (h^{\sst (1)}_{\sst {\it YN}}
   +\Delta m + \overline{v}_{\sst {\it YN}})
   |{\it YN}{\it ;}k_{1} \rangle = E_{k_{1}}|{\it YN}{\it ;}k_{1} \rangle, 
\end{eqnarray}
where we define $h^{\sst (1)}_{\sst {\it YN}}$ as
\begin{eqnarray}
\label{eq:SHYN1}
  h_{\sst {\it YN}}^{\sst (1)}
=
  \left( \begin{array}{cc}
             t_{\sst {\it \Lambda }}
            +u_{\sst {\it \Lambda }}^{\sst (1)}
            +t_{\sst {\it N}}
            +u_{\sst {\it N}} &       0  \\
        0                             & t_{\sst {\it \Sigma }}
                                       +t_{\sst {\it N}}
                                       +u_{\sst {\it N}}
  \end{array} \right).
\end{eqnarray}
The $ u_{\sst {\it \Lambda }}^{\sst (1)} $
is the single-particle potential of $ {\it \Lambda } $ to be calculated
with the $ {\it \Lambda N} $ effective interaction self-consistently.
The number $k_{1}$ stands for a set of quantum numbers to specify a two-body
${\it YN}$ eigenstate. The unitarily transformed ${\it YN}$ interaction in the
first-step calculation is given by
\begin{eqnarray}
\label{eq:VYN1}
  \tilde{v}_{\sst {\it YN}}
           ^{\sst (1)}
  =e^{-S_{\it YN}^{\sst (1)}}
     (h^{\sst (1)}_{\sst {\it YN}}
      +\Delta m+\overline{v}_{\sst {\it YN}})
   e^{ S_{\it YN}^{\sst (1)}}
     -(h_{\sst {\it YN}}^{\sst (1)}
        +\Delta m).
\end{eqnarray}
The decoupling equation for determining the correlation operator
$S_{\sst {\it YN}}^{\sst (1)}$ becomes
\begin{eqnarray}
\label{eq:DECE1}
\overline{Q}_{\sst {\it YN}}^{\sst (1)}
 \tilde{v}_{\sst {\it YN}}^{\sst (1)}
  P_{\sst {\it \Lambda N}}^{\sst (1)}
= P_{\sst {\it \Lambda N}}^{\sst (1)}
 \tilde{v}_{\sst {\it YN}}^{\sst (1)}
    \overline{Q}_{\sst {\it YN}}^{\sst (1)}=0.
\end{eqnarray}
The $ {\it \Lambda N} $ effective interaction
$ \tilde{v}_{\sst {\it YN}}^{\sst (1)} $
in the $ P_{\sst {\it \Lambda N}}^{\sst (1)} $ space is
then given by
\begin{eqnarray}
\label{eq:TVLN1}
\tilde{v}_{\sst {\it \Lambda N}}^{\sst (1)}=
  P_{\sst {\it \Lambda N}}^{\sst (1)}
  \tilde{v}_{\sst {\it YN}}^{\sst (1)}
  P_{\sst {\it \Lambda N}}^{\sst (1)}.
\end{eqnarray}
In the first-step procedure we solve the decoupling equation (\ref{eq:DECE1})
by introducing some assumptions.
As we take a sufficiently large number as $\rho _{\sst 1}$,
we may assume that the one-body potential of ${\it \Lambda}$ for a state
in the  $Q_{\sst {\it \Lambda N}}^{\sst (1)}$ space
 is negligible in comparison with  the sum of the energies,
$t_{\sst {\it \Lambda }}+t_{\sst {\it N}}
+u_{\sst {\it N}}$.
Therefore, $u_{\sst {\it \Lambda }}^{\sst (1)}$ takes
nonzero values only for states in the
$P_{\sst {\it \Lambda N}}^{\sst (1)}$ space,
and we have
\begin{eqnarray}
u_{\sst {\it \Lambda }}^{\sst (1)}=
P_{\sst {\it \Lambda N}}^{\sst (1)}
u_{\sst {\it \Lambda }}^{\sst (1)}
P_{\sst {\it \Lambda N}}^{\sst (1)}.
\end{eqnarray}

The definitions of $ P_{\sst {\it \Lambda N}}^{\sst (1)} $ and
$ Q_{\sst {\it \Lambda N}}^{\sst (1)} $ in Eq.~(\ref{eq:DTPS})
allow us to divide a space of $ {\it \Lambda N} $
states written in relative and center-of-mass (c.m.) coordinates
into the same subspaces
$ P_{\sst {\it \Lambda N}}^{\sst (1)} $ and
$ Q_{\sst {\it \Lambda N}}^{\sst (1)} $ as
\begin{eqnarray}
\label{eq:TPSRC}
|{\it \Lambda N};nlSJ_{r},N_{c}L_{c}\rangle 
  \in \left\{ 
   \begin{array}{ll}
    \displaystyle P_{\sst {\it \Lambda N}}^{\sst (1)} 
      & {\rm if \ }2n+l +2N_{c}+L_{c}\leq \rho _{\sst 1},\\
    \displaystyle Q_{\sst {\it \Lambda N}}^{\sst (1)} 
      & {\rm otherwise},\\
   \end{array}
 \right.
\end{eqnarray}
where $ \{ n,l\} $ and $ \{ N_{c},L_{c} \} $ are the h.o. quantum numbers
of relative and c.m. motions, respectively. The $ S $ and $ J_{r} $
 denote, respectively, spin and angular momentum defined through 
$ \vec{J_{r}}=\vec{l}+\vec{S} $. The isospin $T$ is omitted because
$T$ takes a constant $1/2$ in the $ {\it \Lambda N} $ system.
A $ {\it \Sigma N} $ state in the
$ Q_{\sst {\it \Sigma N}}^{\sst (1)}$ 
space in Fig.~1 is
also written by the product of the h.o. relative and c.m. states.

The $ {\it YN} $ interaction
$ \overline{v}_{\sst {\it YN}} $ in Eq.~(\ref{eq:VBYN}) is not diagonal
in the c.m. quantum numbers $\{ N_{c},L_{c} \}$ because the projection
operators $ P_{\sst {\it \Lambda N}}^{\sst (1)} $,
$\overline{Q}_{\sst {\it YN}}^{\sst (1)}$ and
$Q_{\sst {\it Y_{i}N}}^{\rm \sst (d)}$ are not diagonal
in $\{ N_{c},L_{c} \}$.
Thus, the $ {\it \Lambda N} $ effective interaction
$\tilde{v}_{\sst {\it \Lambda N}}^{\sst (1)}$
is not diagonal in $\{ N_{c},L_{c} \}$.
It has been shown, however, in the calculation of ordinary nuclei that
$\{ N_{c},L_{c} \}$ non-diagonal elements of the $G$-matrix are generally
small \cite{rf:KKKO76}.
Therefore, we make an assumption that the $\{ N_{c},L_{c} \}$ non-diagonal
matrix elements of
$\tilde{v}_{\sst {\it \Lambda N}}^{\sst (1)}$
can be neglected.

The two-body eigenvalue equation (\ref{eq:EIGE1}) is solved for the 
relative motion with a fixed set of $\{ N_{c},L_{c}\} $ written by
\begin{eqnarray}
\label{eq:EIGE1R}
\hspace{-3mm}(h_{\sst {\it YN}}^{\sst (1)}
 + \Delta m+\overline{v}_{\sst {\it YN}})
 |{\it YN};k_{1},ll'SS'J_{r},N_{c}L_{c}\rangle
 = E_{k_{1}}|{\it YN};k_{1},ll'SS'J_{r},N_{c}L_{c}\rangle ,\nonumber \\
\end{eqnarray}
where we rewrite $h_{\sst {\it YN}}^{\sst (1)}$
as
\begin{eqnarray}
\label{eq:SHYN1R}
  h_{\sst {\it YN}}^{\sst (1)}
=
  \left( \begin{array}{cc}
             t_{\sst r{\it \Lambda N}}
            +t_{\sst c{\it \Lambda N}}
            +u_{\sst {\it \Lambda }}^{\sst (1)}
            +u_{\sst {\it N}} &       0  \\
        0                             & t_{\sst r{\it \Sigma N}}
                                       +t_{\sst c{\it \Sigma N}}
                                       +u_{\sst {\it N}}
  \end{array} \right).
\end{eqnarray}
The $t_{\sst rYN} $ and $ t_{\sst cYN}$ are
the kinetic energies of the $ {\it YN} $ relative and c.m. motions, 
respectively.
The state $ |{\it YN};k_{1},ll'SS'J_{r},N_{c}L_{c}\rangle $
expresses a superposition of the $ {\it \Lambda N} $ and $ {\it \Sigma N} $ 
h.o. states and gives a general form of the eigenstate
when we consider the $ {\it \Sigma N} $-$ {\it \Lambda N} $ coupling and the 
${\it YN}$ interaction with the tensor and antisymmetric spin-orbit couplings.
The $ k_{1} $ is an additional quantum number specifying a two-body 
$ {\it YN} $ eigenstate.

In general, the one-body part 
$h_{\sst {\it YN}}^{\sst (1)}+\Delta m$ 
does not satisfies the assumption of decoupling in Eq.~(\ref{eq:ASSUME}).
However, it has been shown in the calculation in nuclei that 
if we take a larger number as $ \rho _{\sst 1}$,
the effect of violating the assumption of decoupling
becomes considerably smaller \cite{rf:SO,rf:KSO}.
We have confirmed that this non-decoupling effect becomes negligible
in calculating nuclear effective interactions.
We here assume this to be the case also
in the determination of the $ {\it \Lambda N} $ effective interaction.
Furthermore, we introduce another approximation that the sum of 
$ t_{\sst c{\it Y_{i}N}}$ and
$ u_{\sst {\it Y_{i}}}+u_{\sst {\it N}}$ be 
diagonal in the c.m. quantum numbers $ N_{c} $ and $ L_{c} $.
This assumption has also been made sure to be acceptable in 
the previous works \cite{rf:SO,rf:KSO,rf:FOS}.

In order to solve the eigenvalue equation (\ref{eq:EIGE1R}) we need to
calculate the matrix elements of
$u_{\sst {\it Y_{i}}}^{\sst (1)}
+u_{\sst {\it N}}$
in the relative and c.m. coordinates representation.
However, the calculation of these matrix elements is difficult
because the term
$u_{\sst {\it Y_{i}}}^{\sst (1)}
+u_{\sst {\it N}}$
can not be separated into a sum of operators written in
relative and c.m. coordinates.
Therefore, as have made in the previous work \cite{rf:SO}, we employ the
angle-average approximation and have
\begin{eqnarray}
\label{eq:UYN}
\lefteqn{\langle {\it Y_{i}N};nlSJ_{r},N_{c}L_{c}|
        u_{\sst {\it Y_{i}}}^{\sst (1)}
       +u_{\sst {\it N}}
          |{\it Y_{i}N};n'l'S'J_{r},N_{c}L_{c}\rangle }\nonumber \\
    &=&\sum_{\stackrel{\scriptstyle n_{\sst {\it Y_{i}}}
                                    l_{\sst {\it Y_{i}}}
                                    n_{\sst {\it N}}
                                    l_{\sst {\it N}}}
                                   {n_{\sst {\it Y_{i}}}'
                                    n_{\sst {\it N}}'\lambda }}
      \frac{2\lambda +1}{(2l+1)(2L_{c}+1)}
        \langle nlN_{c}L_{c}:\lambda 
          |n_{\sst {\it Y_{i}}}
           l_{\sst {\it Y_{i}}}
           n_{\sst {\it N}}
           l_{\sst {\it N}}:\lambda\rangle_{\sst {\it Y_{i}N}}
                                                           \nonumber \\
    &\times& (\ \langle n'l'N_{c}L_{c}:\lambda 
          |n_{\sst {\it Y_{i}}}'
           l_{\sst {\it Y_{i}}}
           n_{\sst {\it N}}
           l_{\sst {\it N}}:\lambda \rangle
               _{\sst {\it Y_{i}N}}
               \langle n_{\sst {\it Y_{i}}}
                       l_{\sst {\it Y_{i}}}|
                 u_{\sst {\it Y_{i}}}^{\sst (1)}
                      |n_{\sst {\it Y_{i}}}'
                       l_{\sst {\it Y_{i}}}\rangle  \nonumber \\
    & &
                  +\langle n'l'N_{c}L_{c}:\lambda 
          |n_{\sst {\it Y_{i}}}
           l_{\sst {\it Y_{i}}}
           n_{\sst {\it N}}'
           l_{\sst {\it N}}:\lambda \rangle
               _{\sst {\it Y_{i}N}}
               \langle n_{\sst {\it N}}
                       l_{\sst {\it N}}\,|
                 u_{\sst {\it N}}
                      |n_{\sst {\it N}}'
                       l_{\sst {\it N}}\,\rangle \ ) 
               \delta _{ll'}\delta _{SS'},
                                    ({\it Y_{i}}={\it \Lambda, \Sigma }),
              \nonumber \\
\end{eqnarray}
where, as was assumed previously, $u_{\sst {\it Y_{i}}}
^{\sst (1)}=0$ for $ {\it Y_{i}} = {\it \Sigma } $.
The $\langle nlN_{c}L_{c}:\lambda |
n_{\sst {\it Y_{i}}}
l_{\sst {\it Y_{i}}}
n_{\sst {\it N}}
l_{\sst {\it N}}:\lambda \rangle _{\sst {\it Y_{i}N}}$ is
the h.o. transformation bracket \cite{rf:BM96} for the ${\it YN}$ two-body
state which 
depends on the rest masses for $ {\it Y_{i}}={\it \Lambda } $ and
$ {\it \Sigma } $.
Here the angle-average approximation means that,
when a function $ f(jm) $ is given with an angular momentum $ j $ and its
$z$-component $m$, we approximate $ f(jm) $ as
$\overline{f}(j)=\sum_{m}f(jm)/(2j+1)$. In the present case the matrix element
on the left-hand side in Eq.~(\ref{eq:UYN}) is dependent on the $z$-components
$ m_{\sst J_{r}} $ and $ m_{\sst L_{c}} $
of $ J_{r} $ and $ L_{c} $, respectively.
The average over
$ m_{\sst J_{r}} $ and $ m_{\sst L_{c}} $
leads to the expression given in Eq.~(\ref{eq:UYN}).
This approximation is essentially the same as has been made to derive
an angle-averaged Pauli operator by Wong and Sauer \cite{rf:Wong,rf:Sauer}.

The use of the basis states $ \{ |{\it Y_{i}N};nlSJ_{r},N_{c}L_{c}\rangle \} $
makes it difficult to represent the operator
$\overline{Q}_{\sst {\it YN}}^{\sst (1)}$,
but it has been
confirmed that the angle-average technique works well.
We thus make an approximation to the operator
$\overline{Q}_{\sst {\it YN}}^{\sst (1)}$
as
\begin{eqnarray}
\label{eq:QBYN1A}
\overline{Q}_{\sst {\it YN}}^{\sst (1)}
\simeq \sum_{\stackrel{\scriptstyle nlN_{c}L_{c}SJ_{r}}{{\it Y_{i}}=
{\it \Lambda, \Sigma}}}
\theta _{\sst {\it Y_{i}N}}(nl,N_{c}L_{c})
|{\it Y_{i}N};nlSJ_{r},
N_{c}L_{c} \rangle \langle {\it Y_{i}N};nlSJ_{r},N_{c}L_{c}|, \nonumber \\
\end{eqnarray}
where the summation should be made on the conditions
\begin{eqnarray}
\label{eq:SUMC1}
& & \rho _{\sst 1}<2n+l+2N_{c}+L_{c}
       \leq \rho _{\rm \sst X}\ {\rm for}\ {\it Y_{i}}
                                                        ={\it \Lambda },
\nonumber \\
& & 0 \leq 2n+l+2N_{c}+L_{c}
\leq \rho _{\rm \sst X}\ {\rm for}\ {\it Y_{i}}={\it \Sigma }.
\end{eqnarray}
Here the coefficient $\theta _{\sst {\it Y_{i}N}}
(nl,N_{c}L_{c})$ 
is defined as
\begin{eqnarray}
\label{eq:THYN}
\lefteqn{\theta _{\sst {\it Y_{i}N}}(nl,N_{c}L_{c})} \nonumber \\
&=& 1-\sum_{n_{\sst {\it Y_{i}}}
l_{\sst {\it Y_{i}}}
n_{\sst {\it N}}l_{\sst {\it N}}\lambda }
\frac{(2\lambda +1)}{(2l+1)(2L_{c}+1)}
\langle nlN_{c}L_{c}:\lambda |n_{\sst {\it Y_{i}}}
l_{\sst {\it Y_{i}}}
n_{\sst {\it N}}l_{\sst {\it N}}:\lambda
\rangle _{\sst {\it Y_{i}N}}^{2},\nonumber \\
& &\hspace{90mm}({\it Y_{i}}={\it \Lambda, \Sigma }),
\end{eqnarray}
where in the summation the following conditions should be satisfied
\begin{eqnarray}
\label{eq:SUMC2}
& & 0 \leq 2n_{\sst {\it N}}+l_{\sst {\it N}}
\leq \rho _{\rm \sst F}, \nonumber \\
& &\rho _{\sst 1}<
 2n_{\sst {\it Y_{i}}}+l_{\sst {\it Y_{i}}}
+2n_{\sst {\it N}}+l_{\sst {\it N}}
\leq \rho _{\rm \sst X} \ {\rm for}\ {\it Y_{i}}=
                                                {\it \Lambda }, \nonumber \\
& & 0 \leq
 2n_{\sst {\it Y_{i}}}+l_{\sst {\it Y_{i}}}
+2n_{\sst {\it N}}+l_{\sst {\it N}}
\leq \rho _{\rm \sst X} \ {\rm for}\ {\it Y_{i}}={\it \Sigma }.
\end{eqnarray}
Furthermore, we need an expression of
$Q^{\rm \sst (d)}_{\sst {\it YN}}
(\nu_{\sst {\it N}})$ given in Eq.~(\ref{eq:QYND})
in the angle-average approximation 
\begin{eqnarray}
\label{eq:QYNDA}
\lefteqn{Q_{\sst {\it Y_{i}N}}^{\rm \sst (d)}
(\nu_{\sst {\it N}})}\nonumber \\
&\simeq & \sum_{\scriptstyle nlN_{c}L_{c}SJ_{r}}
\theta^{'}_{\sst {\it Y_{i}N}}
(nl,N_{c}L_{c},n_{\sst {\it N}} l_{\sst {\it N}})
|{\it Y_{i}N};nlSJ_{r},N_{c}L_{c} \rangle
 \langle {\it {Y_i}N};nlSJ_{r},N_{c}L_{c}|, \nonumber \\
& &\hspace{90mm}({\it Y_{i}}={\it \Lambda, \Sigma }),
\end{eqnarray}
where the summation should be made on the same conditions
as in Eq.~(\ref{eq:SUMC1}). 
The coefficient 
$\theta _{\sst {\it Y_{i}N}}^{'}(nl,N_{c}L_{c},
n_{\sst {\it N}}l_{\sst {\it N}})$ is given by
\begin{eqnarray}
\label{eq:THDYN}
\lefteqn{\theta _{\sst {\it Y_{i}N}}^{'}(nl,N_{c}L_{c},
n_{\sst {\it N}}l_{\sst {\it N}})}\nonumber \\
&=&\sum_{n_{{\it Y_{i}}}l_{{\it Y_{i}}}\lambda }
\frac{(2\lambda +1)}{(2l+1)(2L_{c}+1)}
\langle nlN_{c}L_{c}:\lambda |n_{\sst {\it Y_{i}}}
l_{\sst {\it Y_{i}}}
n_{\sst {\it N}}l_{\sst {\it N}}:\lambda \rangle _{\sst {\it Y_{i}N}}^{2},
({\it Y_{i}}={\it \Lambda, \Sigma }),\nonumber \\
\end{eqnarray}
where $n_{\sst {\it Y_{i}}}$ and $l_{\sst {\it Y_{i}}}$ should
satisfy the conditions given in Eq.~(\ref{eq:SUMC2}).

Once the set of the eigenstates 
$|{\it YN};k_{1},ll'SS'J_{r},N_{c}L_{c}\rangle  $
in Eq.~(\ref{eq:EIGE1R}) is solved,
the $ {\it \Lambda N} $ effective interaction
$\tilde{v}_{\sst {\it \Lambda N}}^{\sst (1)} $ acting
 in the $ P_{\sst {\it \Lambda N}}^{\sst (1)} $ space
 is given according to Eqs.~(\ref{eq:EVLN})-(\ref{eq:NHEVLN}) 
by substituting 
$\omega _{\sst {\it YN}}
 =\omega _{\sst {\it YN}}^{\sst (1)}$, where
$\omega _{\sst {\it YN}}^{\sst (1)}$ is defined as
\begin{eqnarray} 
\label{eq:OYN1}
\hspace{-5mm}\omega _{\sst {\it YN}}^{\sst (1)}
=\sum_{k_{1},ll'SS'J_{r},N_{c}L_{c}}
\overline{Q}_{\sst {\it YN}}^{\sst (1)}
|{\it YN};k_{1},ll'SS'J_{r},N_{c}L_{c}\rangle 
\langle \overline{{\it YN};k_{1},ll'SS'J_{r},N_{c}L_{c}}
|P_{\sst {\it \Lambda N}}^{\sst (1)}.\nonumber \\
\end{eqnarray}
With the $ {\it \Lambda N} $ effective interaction
$\tilde{v}_{\sst {\it \Lambda N}}^{\sst (1)}$
thus determined, we calculate a new single-particle potential
$\langle n_{\sst {\it \Lambda }}
         l_{\sst {\it \Lambda }}|
      u_{\sst {\it \Lambda }}^{\sst (1)}
        |n_{\sst {\it \Lambda }}'
        l_{\sst {\it \Lambda }}\rangle$
as an averaged matrix element of 
$\{ \langle n_{\sst {\it \Lambda }}\\
         l_{\sst {\it \Lambda }}
         j_{\sst {\it \Lambda }}|
      u_{\sst {\it \Lambda }}^{\sst (1)}
        |n_{\sst {\it \Lambda }}'
         l_{\sst {\it \Lambda }}
         j_{\sst {\it \Lambda }}\rangle, 
 j_{\sst {\it \Lambda }}=l_{\sst {\it \Lambda }}\pm1/2 \}$.
The 
$\langle n_{\sst {\it \Lambda }}
         l_{\sst {\it \Lambda }}|
      u_{\sst {\it \Lambda }}^{\sst (1)}
        |n_{\sst {\it \Lambda }}'
        l_{\sst {\it \Lambda }}\rangle$
can be given explicitly in terms of the effective interaction
$\tilde{v}_{\sst {\it \Lambda N}}^{\sst (1)}$
as
\begin{eqnarray}
\label{eq:UL1}
\langle n_{\sst {\it \Lambda }}l_{\sst {\it \Lambda }}|
u_{\sst {\it \Lambda }}^{\sst (1)}
|n_{\sst {\it \Lambda }}'l_{\sst {\it \Lambda }}\rangle
&\equiv & 
\frac{\sum_{j_{\sst {\it \Lambda }}}
 (2j_{\sst {\it \Lambda }}+1)\langle n_{\sst {\it \Lambda }}
l_{\sst {\it \Lambda }}j_{\sst {\it \Lambda }}
|u_{\sst {\it \Lambda }}^{\sst (1)}
|n_{\sst {\it \Lambda }}'l_{\sst {\it \Lambda }}
j_{\sst {\it \Lambda }}\rangle }{\sum_{j_{\sst {\it \Lambda }}
=l_{\sst {\it \Lambda }}\pm \frac{1}{2}}(2j_{\sst {\it \Lambda }}+1)}
\nonumber \\
&=& \sum_{\stackrel {\scriptstyle nn'lSJ_{r}N_{c}L_{c}\lambda }
{\{ n_{\sst {\it N}},l_{\sst {\it N}}\} :{\rm occupied}}}
 \frac{(2\lambda +1)(2J_{r}+1)}{(2l_{\sst {\it \Lambda }}+1)(2l+1)}
\nonumber \\
&\times & \langle nlN_{c}L_{c}:\lambda |n_{\sst {\it \Lambda }}
l_{\sst {\it \Lambda }}n_{\sst {\it N}}l_{\sst {\it N}}:
\lambda \rangle _{\sst {\it \Lambda N}}
    \langle n'lN_{c}L_{c}:\lambda |n_{\sst {\it \Lambda }}'
l_{\sst {\it \Lambda }}n_{\sst {\it N}}l_{\sst {\it N}}:
\lambda \rangle _{\sst {\it \Lambda N}} \nonumber \\
&\times & \langle {\it \Lambda N};nlSJ_{r},N_{c}L_{c}
|\tilde{v}_{\sst {\it \Lambda N}}^{\sst (1)}
|{\it \Lambda N};n'lSJ_{r},N_{c}L_{c}\rangle .
\label{eq:17}
\end{eqnarray}
In the above treatment of the single-particle potential of
$ {\it \Lambda } $ in the first-step
calculation, we do not consider the one-body spin-orbit splitting.

The $ {\it \Lambda N} $ effective interaction
$ \tilde{v}_{\sst {\it \Lambda N}}^{\sst (1)} $
in Eq.~(\ref{eq:TVLN1}) is dependent on the single-particle potential
$ u_{\sst {\it \Lambda }}^{\sst (1)} $,
and $ u_{\sst {\it \Lambda }}^{\sst (1)} $
is determined with
$ \tilde{v}_{\sst {\it \Lambda N}}^{\sst (1)} $
as in Eq.~(\ref{eq:UL1}).
Therefore, the equation for determining
$ \tilde{v}_{\sst {\it \Lambda N}}^{\sst (1)} $
and $ u_{\sst {\it \Lambda }}^{\sst (1)} $ should be
solved simultaneously.

\subsection{Second-step decoupling}

The model space $P_{\sst {\it \Lambda N}}^{\sst (1)}$ 
in the first-step procedure is chosen as a {\it low-momentum} space,
but the $P_{\sst {\it \Lambda N}}^{\sst (1)}$ 
space is still large when we wish to use
$ \tilde{v}_{\sst {\it \Lambda N}}^{\sst (1)} $
in the calculation of bound or low
excited states of $ {\it \Lambda } $ hypernuclei.
With $ \tilde{v}_{\sst {\it \Lambda N}}^{\sst (1)} $,
we further proceed to the
calculation of the $ {\it \Lambda N} $ effective interaction acting
in a smaller model space and correspondingly the single-particle
potential of $ {\it \Lambda } $ in low-lying states.
We separate the $P_{\sst {\it \Lambda N}}^{\sst (1)}$
 space into the
$ P_{\sst {\it \Lambda N}}^{\sst (2)}$,
$ P_{\sst {\it \Lambda N}}^{\rm \sst (X)}$,
$P_{\sst {\it \Lambda N}}^{\rm \sst (Y)}$
and $Q_{\sst {\it \Lambda N}}^{\sst (2)}$ spaces
specified by the numbers $\rho _{\sst 1}$, $\rho _{\rm \sst F}$,
and $\rho _{\sst 2}$ as shown in Fig.~2.
The number $ \rho _{\sst 2} $ is introduced to specify
the uppermost bound state of $ {\it \Lambda } $.
Here these spaces are defined as 
\begin{eqnarray}
\lefteqn{|\alpha _{\sst {\it \Lambda }}
\beta _{\sst {\it N}}\rangle }\nonumber \\
&\in &\left\{ 
   \begin{array}{ll}
    \displaystyle 
      P_{\sst {\it \Lambda N}}^{\sst (2)}
         & {\rm if \ } 2n_{\sst {\it \Lambda }}
                       +l_{\sst {\it \Lambda }}
                            \leq \rho _{\sst 2}
          \ {\rm and \ }  2n_{\sst {\it N}}
                       +l_{\sst {\it N}} 
                            \leq \rho _{\rm \sst F}, \\
     P_{\sst {\it \Lambda N}}^{\rm \sst (X)}
         & {\rm if \ } 2n_{\sst {\it \Lambda }}
                       +l_{\sst {\it \Lambda }}
                            > \rho _{\sst 2},\ 
                       2n_{\sst {\it N}}
                       +l_{\sst {\it N}} 
                            \leq \rho _{\rm \sst F}
          \ {\rm and \ } 2n_{\sst {\it \Lambda }}
                       +l_{\sst {\it \Lambda }}
                       +2n_{\sst {\it N}}
                       + l_{\sst {\it N}} 
                            \leq \rho _{\sst 1}, \\
     P_{\sst {\it \Lambda N}}^{\rm \sst (Y)}
         & {\rm if \ } 2n_{\sst {\it \Lambda }}
                       +l_{\sst {\it \Lambda }}
                            \leq \rho _{\sst 2},\ 
                       2n_{\sst {\it N}}
                       +l_{\sst {\it N}} 
                            > \rho _{\rm \sst F}
          \ {\rm and \ } 2n_{\sst {\it \Lambda }}
                       +l_{\sst {\it \Lambda }}
                       +2n_{\sst {\it N}}
                       + l_{\sst {\it N}} 
                            \leq \rho _{\sst 1}, \\
     Q_{\sst {\it \Lambda N}}^{\sst (2)}
         & {\rm if \ } 2n_{\sst {\it \Lambda }}
                       +l_{\sst {\it \Lambda }}
                            > \rho _{\sst 2},\ 
                       2n_{\sst {\it N}}
                       +l_{\sst {\it N}} 
                            > \rho _{\rm \sst F}
          \ {\rm and \ } 2n_{\sst {\it \Lambda }}
                       +l_{\sst {\it \Lambda }}
                       +2n_{\sst {\it N}}
                       + l_{\sst {\it N}} 
                            \leq \rho _{\sst 1}. \\
   \end{array}\right.\nonumber \\
\end{eqnarray}

The $P_{\sst {\it \Lambda N}}^{\rm \sst (X)}$ space 
is the excluded space due to the Pauli
 principle for nucleons. 
We are principally interested in the $ {\it \Lambda N} $ effective interaction
that gives a description of the bound states of ${\it \Lambda}$. 
It would be desirable that such a $ {\it \Lambda N} $ effective interaction
satisfies the condition of decoupling between two $ {\it \Lambda N} $ states
consisting of the bound and unbound states of ${\it \Lambda}$.
This requirement is equivalent to the situation that
the $ {\it \Lambda N} $ effective interactions
should be decoupled between two $ {\it \Lambda N} $ states in the 
$P_{\sst {\it \Lambda N}}^{\sst (2)}$
and 
$Q_{\sst {\it \Lambda N}}^{\sst (2)}$
spaces.
Therefore, we solve the decoupling equation for
${\tilde v}^{\sst (2)}_{\sst {\it \Lambda N}}$ 
between the $ {\it \Lambda N} $ states in the 
$P_{\sst {\it \Lambda N}}^{\sst (2)}$
and 
$Q_{\sst {\it \Lambda N}}^{\sst (2)}$
spaces. 

The space $P_{\sst {\it \Lambda N}}^{\rm \sst (Y)}$
is not a space to be excluded due to the Pauli principle.
However, some states in the
$P_{\sst {\it \Lambda N}}^{\rm \sst (Y)}$ and
$P_{\sst {\it \Lambda N}}^{\sst (2)}$ spaces
overlap each other in energy.
Therefore, we treat separately the effect of the coupling between the
$P_{\sst {\it \Lambda N}}^{\rm \sst (Y)}$ and
$P_{\sst {\it \Lambda N}}^{\sst (2)}$ spaces.
Actually the interaction between states in the
$P_{\sst {\it \Lambda N}}^{\rm \sst (Y)}$ and
$P_{\sst {\it \Lambda N}}^{\sst (2)}$ spaces
induces the core polarization in the core nucleus, as discussed in
Refs.~\cite{rf:SO} and \cite{rf:FOS}.
We take into account the effect of the core polarization in the usual
perturbative calculation.

\begin{figure}[t]
  \epsfxsize = 6cm
  \centerline{\epsfbox{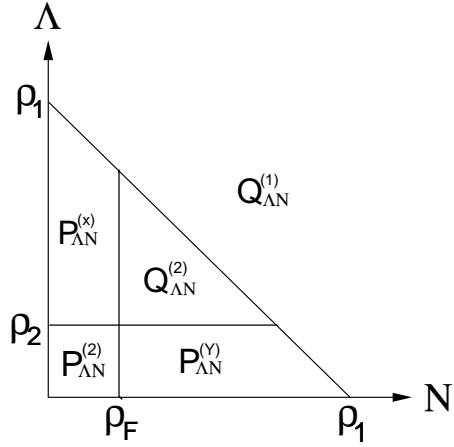}}
  \caption {The model space and its complement in the
 second-step calculation.}
  \label{fig:2}
\end{figure}

In the second-step calculation the unitarily transformed interaction 
is given by
\begin{eqnarray}
\label{eq:TVYN2}
\tilde{v}_{\sst {\it \Lambda N}}^{\sst (2)}
   = e^{-S_{\it \Lambda N}^{(2)}}
  (h_{\sst {\it \Lambda N}}^{\sst (2)}
   +\tilde{v}_{\sst {\it \Lambda N}}^{\sst (1)})
     e^{S_{\it \Lambda N}^{(2)}}
  -h_{\sst {\it \Lambda N}}^{\sst (2)},
\end{eqnarray}
where
\begin{eqnarray}
\label{eq:HLN2}
h_{\sst {\it \Lambda N}}^{\sst (2)}=t_{\sst {\it \Lambda }}
+u_{\sst {\it \Lambda }}^{\sst (2)}+t_{\sst {\it N}}+u_{\sst {\it N}}.
\end{eqnarray}
The term $ u_{\sst {\it \Lambda }}^{\sst (2)} $ 
is the self-consistent potential of $ {\it \Lambda } $ to be determined in the 
second-step procedure.
The decoupling equation for determining the correlation operator
$S_{\sst {\it \Lambda N}}^{\sst (2)}$ in Eq.~(\ref{eq:TVYN2}) becomes
\begin{eqnarray}
\label{eq:DEC2}
Q_{\sst {\it \Lambda N}}^{\sst (2)}
\tilde{v}_{\sst {\it \Lambda N}}^{\sst (2)}
P_{\sst {\it \Lambda N}}^{\sst (2)}
=P_{\sst {\it \Lambda N}}^{\sst (2)}
\tilde{v}_{\sst {\it \Lambda N}}^{\sst (2)}
Q_{\sst {\it \Lambda N}}^{\sst (2)}=0.
\end{eqnarray}
We note here that the assumption of decoupling
\begin{eqnarray}
\label{eq:ASSUME2}
Q_{\sst {\it \Lambda N}}^{\sst (2)}
h_{\sst {\it \Lambda N}}^{\sst (2)}
P_{\sst {\it \Lambda N}}^{\sst (2)}=0
\end{eqnarray}
is satisfied exactly because the two two-body states in the 
$P_{\sst {\it \Lambda N}}^{\sst (2)}$
and $Q_{\sst {\it \Lambda N}}^{\sst (2)}$ spaces
do not contain a common single-particle state of ${\it \Lambda}$
or $ {\it N} $.

The two-body eigenvalue equation to be solved in the second-step 
calculation is written as
\begin{eqnarray}
\label{eq:EIGE2}
 (P_{\sst {\it \Lambda N}}^{\sst (2)}
 +Q_{\sst {\it \Lambda N}}^{\sst (2)} )
   (h_{\sst {\it \Lambda N}}^{\sst (2)}
   +\tilde{v}_{\sst {\it \Lambda N}}^{\sst (1)})
 (P_{\sst {\it \Lambda N}}^{\sst (2)}
 +Q_{\sst {\it \Lambda N}}^{\sst (2)} )
   |{\it \Lambda N};k_{2},J\rangle
    = E_{k_{2}}|{\it \Lambda N};k_{2},J\rangle ,
\end{eqnarray}
where $J$ denotes the total angular momentum, 
and $k_{2}$ a set of quantum numbers of a $ {\it \Lambda N} $ two-body
eigenstate.
It should be noted that this eigenvalue equation can be solved exactly in the
$ P_{\sst {\it \Lambda N}}^{\sst (2)}
+Q_{\sst {\it \Lambda N}}^{\sst (2)} $ space by the 
diagonalization of the hamiltonian written with the
basis states 
$\{|n_{\sst {\it \Lambda }}
    l_{\sst {\it \Lambda }}
    j_{\sst {\it \Lambda }},
    n_{\sst {\it N}}
    l_{\sst {\it N}}
    j_{\sst {\it N}};JM\rangle \}$ if the 
 $P_{\sst {\it \Lambda N}}^{\sst (2)}
 +Q_{\sst {\it \Lambda N}}^{\sst (2)}$ space
is chosen suitably as a small space.
Once a set of the eigenstates $|{\it \Lambda N};k_{2},J\rangle$ 
in Eq.~(\ref{eq:EIGE2}) is solved,
the $ {\it \Lambda N} $ effective interaction 
$\tilde{v}_{\sst {\it \Lambda N}}^{\sst (2)}$
 in the 
$P_{\sst {\it \Lambda N}}^{\sst (2)}$ space
 is given from Eqs.~(\ref{eq:EVLN})-(\ref{eq:NHEVLN})
with 
 $\omega _{\sst {\it YN}}
 =\omega _{\sst {\it \Lambda N}}^{\sst (2)} $ defined by
\begin{eqnarray}
\label{eq:OYN2}
\omega _{\sst {\it \Lambda N}}^{\sst (2)}
=\sum_{k_{2},J} Q_{\sst {\it \Lambda N}}^{\sst (2)}
|{\it \Lambda N};k_{2},J\rangle \langle \overline{{\it \Lambda N};k_{2},J}
|P_{\sst {\it \Lambda N}}^{\sst (2)}.
\end{eqnarray}

The single-particle potential
$u_{\sst {\it \Lambda }}^{\sst (2)} $ 
is calculated with the $ {\it \Lambda N} $ effective interactions,
$\tilde{v}_{\sst {\it \Lambda N}}^{\sst (2)}$ and
$\tilde{v}_{\sst {\it \Lambda N}}^{\sst (1)}$, as
\begin{eqnarray}
\label{eq:UL2}
  \lefteqn{\langle \alpha_{\sst {\it \Lambda }}|
   u_{\sst {\it \Lambda }}^{\sst (2)}
         |\alpha_{\sst {\it \Lambda }}'\rangle }\nonumber \\
&=&\sum_{\beta_{\sst {\it N}}:{\rm occupied}}
  \left\{ 
   \begin{array}{ll}
         \langle \alpha_{\sst {\it \Lambda }}
                  \beta_{\sst {\it N}}|
        \tilde{v}_{\sst {\it \Lambda N}}^{\sst (2)}
                |\alpha_{\sst {\it \Lambda }}'
                  \beta_{\sst {\it N}} \rangle
                    &    {\rm if \ }
                          |\alpha_{\sst {\it \Lambda }}
                           \beta_{\sst {\it N}}\rangle
             \in P_{\sst {\it \Lambda N}}^{\sst (2)}
                    \ {\rm and}\ 
                          |\alpha_{\sst {\it \Lambda }}'
                           \beta_{\sst {\it N}}\rangle       
             \in P_{\sst {\it \Lambda N}}^{\sst (2)},\\
 \langle \alpha_{\sst {\it \Lambda }}
                  \beta_{\sst {\it N}}|
        \tilde{v}_{\sst {\it \Lambda N}}^{\sst (1)}
                |\alpha_{\sst {\it \Lambda }}'
                  \beta_{\sst {\it N}} \rangle
                    &  {\rm otherwise. \ }   
   \end{array}
  \right.
\end{eqnarray}
We solve the set of equations for determining
$\tilde{v}_{\sst {\it \Lambda N}}^{\sst (2)}$ 
and $u_{\sst {\it \Lambda }}^{\sst (2)}$
 iteratively until the result converges.

We remark here that we employ 
$\tilde{v}_{\sst {\it \Lambda N}}^{\sst (1)}$
as the effective interaction in the 
$P_{\sst {\it \Lambda N}}^{\rm \sst (X)}$
space in the calculation of 
$u_{\sst {\it \Lambda }}^{\sst (2)}$.
Since the effective interaction in the 
$P_{\sst {\it \Lambda N}}^{\rm \sst (X)}$
space is unchanged in the transformation made in the second-step
procedure, we may write the $ {\it \Lambda N} $ effective interaction
in the 
$P_{\sst {\it \Lambda N}}^{\rm \sst (X)}$
space as 
$P_{\sst {\it \Lambda N}}^{\rm \sst (X)}
 \tilde{v}_{\sst {\it \Lambda N}}^{\sst (2)}
 P_{\sst {\it \Lambda N}}^{\rm \sst (X)}
=P_{\sst {\it \Lambda N}}^{\rm \sst (X)}
 \tilde{v}_{\sst {\it \Lambda N}}^{\sst (1)}
 P_{\sst {\it \Lambda N}}^{\rm \sst (X)}$.
We, therefore, have used partly the $ {\it \Lambda N}$ effective interaction
$\tilde{v}_{\sst {\it \Lambda N}}^{\sst (1)}$
 in the calculation of 
$u_{\sst {\it \Lambda }}^{\sst (2)}$
 as in Eq.~(\ref{eq:UL2}).

\subsection{Corrections in perturbative treatment}

First we note that the one-body hamiltonian
$h_{\sst {\it \Lambda N}}^{\sst (2)}$ 
in Eq.~(\ref{eq:HLN2}) has non-diagonal
terms because we did not make the Hartree-Fock calculation. 
These terms could bring about sizable corrections to the
single-particle energies of $\Lambda$.  We adopt the same
h.o. energy both for ${\it \Lambda}$ and $ {\it N} $, though it has been
considered that the wave function of ${\it \Lambda}$ is basically
 different from that of $ {\it N}$. Therefore, we must evaluate the
 corrections arising from the non-diagonal terms of 
$h_{\sst {\it \Lambda N}}^{\sst (2)}$. 
By taking the diagonal part of 
$h_{\sst {\it \Lambda N}}^{\sst (2)}$
as the unperturbed hamiltonian we calculate these corrections
perturbatively up to second order as shown in diagram (a)
of Fig.~3.

Next, the interaction between the $ {\it \Lambda N} $ states in the
$P_{\sst {\it \Lambda N}}^{\sst (2)}$
and 
$P_{\sst {\it \Lambda N}}^{\rm \sst (Y)}$
spaces induces the core polarization in the core nucleus,
 and is important in the calculation of the corrections. 
We here assume that these corrections can be calculated
 approximately, using the interaction
$P_{\sst {\it \Lambda N}}^{\sst (2)}
 \tilde{v}_{\sst {\it \Lambda N}}^{\sst (1)}
 P_{\sst {\it \Lambda N}}^{\rm \sst (Y)}
+{\rm h.c.}.$
It will be desirable to take into account the effect of
intermediate nucleon states with high momentum. In this 
perturbative calculation we include only the nucleon states
belonging to the
$P_{\sst {\it \Lambda N}}^{\rm \sst (Y)}$
 space, that is, the nucleon h.o. state should satisfy
$2n_{\sst {\it N}}+l_{\sst {\it N}}\leq
\rho_{\sst 1}
-(2n_{\sst {\it \Lambda }}+l_{\sst {\it \Lambda }})$,
 where 
$\{ n_{\sst {\it \Lambda }},l_{\sst {\it \Lambda }}\}$
are the h.o. quantum numbers of a ${\it \Lambda}$ state.
The validity
of this truncation of nucleon states will
be checked  in the numerical calculation in the next section
by increasing the number $\rho_{\sst 1}$.
The diagrams corresponding these corrections are given in
diagram (b) and (c) of Fig.~3.

\begin{figure}[t]
  \epsfxsize = 9.6cm
  \centerline{\epsfbox{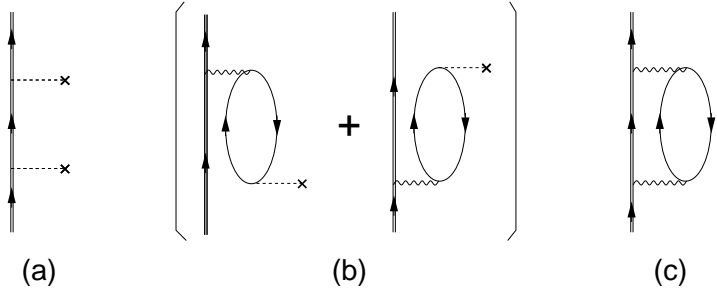}}
  \caption {Calculated diagrams for the ${\it \Lambda}$
single-particle energies.
 The $ {\it \Lambda N} $ effective interaction
$\tilde{v}_{\sst {\it \Lambda N}}$ is
 represented by the wavy-line vertex, and the $\times $ vertex with the 
 dashed line represents the non-diagonal part of the one-body hamiltonian
$ h_{k}=t_{k}+u_{k} $ for $ k={\it \Lambda } $ and $ {\it N} $.
The propagations of $ {\it N} $ and $ {\it \Lambda } $ are represented
by the single and double external lines, respectively.}
  \label{fig:3}
\end{figure}

\section{Application to $_{\it \Lambda }^{17}$O}
\setcounter{equation}{0}

We performed calculations employing the J\"ulich
model-\~A (J\~A), model-\~B (J\~B) \cite{rf:JUL}
\footnote{We used the computer code for the J\"ulich ${\it YN}$
potentials in the momentum-space
representation made by B.~Holzenkamp and A.~Reuber.}
and Nijmegen soft-core (NSC) \cite{rf:NSC}
\footnote{We used the computer code for the Nijmegen soft-core
{\it YN} potential in the coordinate-space representation provided by
T.~A.~Rijken and J.~J.~de~Swart.}
potentials  
 for the {\it YN} interaction. We used the h.o. basis states with a common
 $ {\it \hbar \Omega }=14 $MeV for $ {\it \Lambda } $, $ {\it \Sigma } $
and $ {\it N} $. 
The dependence of the effective interaction and single-particle
energies on the value
$ {\it \hbar \Omega } $ has been already discussed in $^{16}$O \cite{rf:SO86}
and we confirmed that the dependence was quite small
around ${\it \hbar \Omega }=14 $MeV if perturbation corrections were included.
As for the nucleon single-particle potential $ u_{\sst {\it N}} $,
we used the fixed
data \cite{rf:SO} calculated using the Paris potential \cite{rf:Paris},
and we have assumed that $ u_{\sst {\it \Sigma }}=0 $.
The boundary numbers $ \rho _{\sst 2}$,
$ \rho _{\rm \sst F}$ and $ \rho _{\rm \sst X} $
are taken as
 $ \rho _{\sst 2}=1 $, $ \rho _{\rm \sst F}=1 $
and $ \rho _{\rm \sst X}=12 $ for $_{\it \Lambda }^{17}$O.

\begin{figure}[t]
  \epsfxsize = 13.8cm
  \centerline{\epsfbox{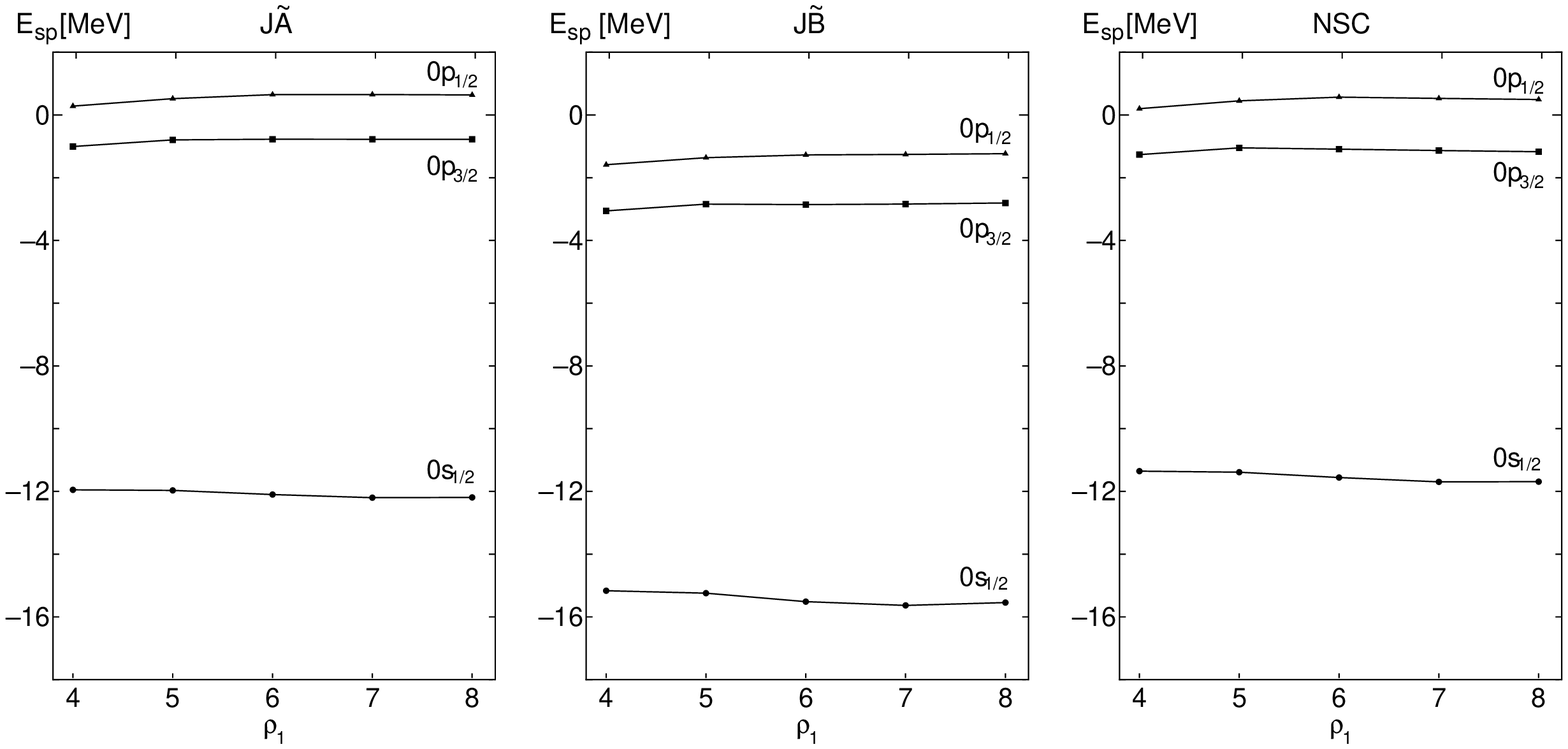}}
  \caption {The $ \rho _{\sst 1} $ dependence of the calculated
$ {\it \Lambda } $
 single-particle energies including the perturbation corrections
given in Fig.~3.}
  \label{fig:4}
\end{figure}

We present, in Fig.~4, the $ \rho _{\sst 1} $ dependence of
the single-particle energies of $ {\it \Lambda } $ including the
perturbation corrections shown in Fig.~3.
The parameter $ \rho _{\sst 1} $ specifies
the $P_{\sst {\it \Lambda N}}^{\sst (1)}$ space
in which the $ {\it \Lambda N} $ effective interaction
$\tilde{v}_{\sst {\it \Lambda N}}^{\sst (2)}$ is
determined rigorously without any approximation as discussed in  Sec.~3.
Therefore, as $\rho _{\sst 1}$ becomes larger, the
calculated result does more accurate. If the result converges at a
certain number $ \rho_{\sst 1} $, we may say that the method of
the two-step calculation of the $ {\it \Lambda N} $ effective
interaction would be trustworthy.
As shown in Fig.~4, the results are convergent and stable for
$ \rho _{\sst 1}\geq 7$ for each ${\it YN}$ potential.
If we take $ \rho _{\sst 1} $ larger than 6,
our model space is considered to be sufficiently large.
We note that
the model space with $\rho _{\sst 1}=6$ corresponds to that
used in the $G$-matrix calculations made by Halderson \cite{rf:HAL93} and by
Hao and Kuo \cite{rf:HAO93,rf:HAO96}.

\begin{table}[b]
\caption{Dependence of typical diagonal matrix elements
$\langle ab|\tilde{v}_{\sst {\it \Lambda N}}^{\sst (2)}|ab\rangle _J$ on
$N_{\rm max}$ in units of MeV,
where $ a $ and $ b $ denote the $ {\it \Lambda } $
 and $ {\it N} $ single-particle orbits, respectively,
labeled as $ 1=0s_{1/2} $,
 $ 2=0p_{1/2} $ and $ 3=0p_{3/2} $, and $J$ is the total angular
 momentum.}
\label{table:1}
  \begin{center}
    \begin{tabular}{cccccccc} \hline
          & $a=2,$ & $b=2,$ & $J=1$ & & $a=3,$ & $b=2,$ & $J=2$  \\
\cline{2-4} \cline{6-8}
$N_{\rm max}$ &J\~A&  J\~B  & NSC    & & J\~A   & J\~B   & NSC \\
\cline{2-8}
      40  & -0.991 & -1.331 & -0.347 & &-0.825 & -0.828 & -1.146 \\
      60  & -1.100 & -1.432 & -0.402 & &-0.909 & -0.892 & -1.186 \\
      80  & -1.145 & -1.483 & -0.408 & &-0.932 & -0.920 & -1.191 \\
     100  & -1.152 & -1.492 & -0.412 & &-0.937 & -0.926 & -1.195 \\
     120  & -1.152 & -1.492 & -0.417 & &-0.937 & -0.925 & -1.198 \\
     140  &        &        & -0.422 & &       &        & -1.201 \\
     160  &        &        & -0.426 & &       &        & -1.204 \\
     180  &        &        & -0.428 & &       &        & -1.205 \\
     200  &        &        & -0.428 & &       &        & -1.205 \\ \hline
    \end{tabular}
  \end{center}
\end{table}

Table~1 shows the dependence of typical matrix elements of the
$ {\it \Lambda N} $ effective interaction at $ \rho _{\sst 1}=8 $ on
the value $N_{\rm max}$ which denotes the number of the h.o. basis functions.
The number $N_{\rm max}$ determines the dimension of the space of
relative states in the first-step calculation.
When we solve the eigenvalue equation (\ref{eq:EIGE1R}), taking into account
the tensor and $ {\it \Sigma N} $-$ {\it \Lambda N} $ couplings, we need
2$\times$2$\times$$N_{\rm max}$ basis functions in each partial-wave channel.
Table~1 shows that the results are convergent if we take 120 for the
J\~A and J\~B potentials, and 200 for the NSC potential
as $N_{\rm max}$.
We have also made sure that the situation of convergence is similar in the
other matrix elements of the $ {\it \Lambda N} $ effective interaction.
As we employ the NSC potential given in the configuration-space representation,
we need much more basis states than in the J\"ulich potential given in the
momentum-space representation in order to treat short-range correlations
accurately.

\begin{table}[b]
\caption{Diagonal matrix elements
 $\langle ab|\tilde{v}_{\sst {\it \Lambda N}}^{\sst (2)}|ab\rangle _J$
in units of MeV. The notations are the same in Table 1.}
  \label{table:2}
  \begin{center}
    \begin{tabular}{ccccccccccccc} \hline
 $a$&$b$&$J$& J\~A & J\~B & NSC &  &$a$&$b$&$J$& J\~A & J\~B & NSC \\
\cline{1-6}\cline{8-13}
  1 & 1 & 0 & -1.31 & -0.06 & -4.04 &  & 2 & 3 & 1 & -0.38 & -0.75 & -0.42 \\
  1 & 1 & 1 & -2.49 & -3.47 & -1.38 &  & 2 & 3 & 2 & -0.94 & -0.92 & -1.26 \\
  1 & 2 & 0 & -0.78 & -1.22 & -0.39 &  & 3 & 1 & 1 & -0.94 & -0.62 & -1.45 \\
  1 & 2 & 1 & -0.95 & -0.98 & -0.70 &  & 3 & 1 & 2 & -1.38 & -1.80 & -1.06 \\
  1 & 3 & 1 & -0.46 & -0.11 & -1.13 &  & 3 & 2 & 1 & -0.50 & -0.90 & -0.34 \\
  1 & 3 & 2 & -1.17 & -1.50 & -1.00 &  & 3 & 2 & 2 & -0.94 & -0.93 & -1.21 \\
  2 & 1 & 0 & -1.05 & -1.56 & -0.54 &  & 3 & 3 & 0 & -0.49 &  0.65 & -2.03 \\
  2 & 1 & 1 & -0.84 & -0.84 & -0.95 &  & 3 & 3 & 1 & -0.73 & -1.16 & -0.02 \\
  2 & 2 & 0 & -0.10 &  0.57 & -0.71 &  & 3 & 3 & 2 & -0.20 &  0.12 & -0.96 \\
  2 & 2 & 1 & -1.15 & -1.49 & -0.43 &  & 3 & 3 & 3 & -1.34 & -1.78 & -0.84 \\
\hline
    \end{tabular}
  \end{center}
\end{table}

We show, in Table~2, the diagonal matrix elements of the 0$s$-0$p$ shell
$ {\it \Lambda N} $ effective interaction calculated in the second-step
procedure.
Because of the different characteristics of the original $ {\it YN} $
potentials, the results depend strongly on the ${\it YN}$ potentials
used in the calculation.
Considering the situations of convergence shown in Fig.~4 and Table~1, we have
used the numbers $ \rho_{\sst 1}=8 $, and $N_{\rm max}=120$
and $200$ for the J\"ulich and Nijmegen potentials, respectively, in Table~2.
We shall use hereafter the same set of the numbers in the following tables.

In the present calculation we consider three kinds of the treatments of
the Pauli-blocking effect in the first-step procedure.

(1) The Pauli blocking in the spaces
$ Q_{\sst {\it \Lambda N}}^{\rm \sst (X)} $ and
$ Q_{\sst {\it \Sigma N}}^{\rm \sst (X)} $
is not considered, that is, the original $ {\it YN} $ interaction
$ v_{\sst {\it YN}} $ is used in place of the interaction 
$ \overline{v}_{\sst {\it YN}} $ in Eq.~(\ref{eq:VBYN}).

(2) The Pauli blocking in both the spaces
$ Q_{\sst {\it \Lambda N}}^{\rm \sst (X)} $ and
$ Q_{\sst {\it \Sigma N}}^{\rm \sst (X)} $ is considered,
that is, we take only the first term of
$ \overline{v}_{\sst {\it YN}} $ in Eq.~(\ref{eq:VBYN}).

(3) The Pauli blocking in the spaces
$ Q_{\sst {\it \Lambda N}}^{\rm \sst (X)} $ and
$ Q_{\sst {\it \Sigma N}}^{\rm \sst (X)} $ is considered,
but the diagonal interaction, the second term of
$ \overline{v}_{\sst {\it YN}} $ in Eq.~(\ref{eq:VBYN}),
is restored.

It is noted that, in the above three cases, the predominant Pauli-blocking
effect which comes from the low-momentum $ {\it \Lambda N} $ states
contained in the $ P_{\sst {\it \Lambda N}}^{\rm \sst (X)} $ space is
taken into account rigorously in the second-step calculation.

\begin{table}[b]
\caption{Dependence of the single-particle energies of $ {\it \Lambda } $
on the different treatments of the Pauli-blocking effects
for the 0$s_{1/2}$, 0$p_{3/2}$ and 0$p_{1/2}$ states including the
perturbation corrections.}
\label{table:3}
  \begin{center}
    \begin{tabular}{cccccccccccc} \hline
        & & 0$s_{1/2} $ &  & &  & 0$p_{3/2}$ &  & & &  0$p_{1/2}$ &  \\ 
      Pauli  & J\~A & J\~B &  NSC  && J\~A & J\~B & NSC && J\~A & J\~B & NSC \\
\cline{2-4}\cline{6-8}\cline{10-12}
(1)&-14.09 &-16.47 & -16.74 && -2.49 & -3.95 & -4.47 && -1.29 & -2.46 &-3.07 \\
(2)&-10.96 &-13.42 &  -9.23 && -0.16 & -1.76 & -0.04 && 1.28 & -0.17 & 1.68 \\
(3)&-12.19 &-15.54 & -11.69 && -0.78 & -2.81 & -1.17 && 0.64 & -1.24 & 0.49 \\
\hline
    \end{tabular}
  \end{center}
\end{table}
\begin{table}[b]
\caption{
Partial wave contributions to the first-order potential energy of 
$ {\it \Lambda } $ in the 0$s_{1/2}$ state. All entries are in MeV.}
  \label{table:4}
  \begin{center}
    \begin{tabular}{cccccccc} \hline
     & & $^{1}S_{0}$ & $^{3}S_{1}$-$^{3}D_{1}$ & $^{3}P_{0}$ &
      $^{1}P_{1}$-$^{3}P_{1}$ & $^{3}P_{2}$-$^{3}F_{2}$ & Total\\ \hline
                  & (1) & -4.87 & -17.94 & 0.32 & 0.91 & -0.10 & -21.67 \\
  J\~A      & (2) & -2.51 & -17.26 & 0.40 & 1.14 & -0.09 & -18.32 \\
                  & (3) & -3.14 & -17.82 & 0.37 & 1.07 & -0.10 & -19.61 \\
\hline
                  & (1) & -0.24 & -24.81 & 0.31 & 1.43 & 0.31 & -23.01 \\
  J\~B      & (2) &  0.12 & -22.61 & 0.37 & 1.59 & 0.32 & -20.22 \\
                  & (3) & -0.17 & -24.43 & 0.35 & 1.54 & 0.31 & -22.40 \\
\hline
                  & (1) & -9.70 & -14.16 & 0.18 & 1.80 & -1.86 & -23.73 \\
              NSC & (2) & -9.58 & -7.29  & 0.28 & 2.05 & -1.84 & -16.38 \\
                  & (3) & -9.59 & -9.79  & 0.24 & 1.96 & -1.85 & -19.04 \\
\hline
    \end{tabular}
  \end{center}
\end{table}
We here discuss the dependence of the different treatments of the
Pauli-blocking effect in the first-step procedure for three
${\it YN}$ interactions.
In Table~3 we display the single-particle energies of $ {\it \Lambda } $
including the perturbation corrections shown in Fig.~3.
Rows (1), (2) and (3) correspond to different treatments (1), 
(2) and (3) of the
Pauli-blocking effect in the first-step procedure.
All the single-particle energies of $ {\it \Lambda } $ in the 0$s_{1/2}$,
0$p_{3/2}$ and 0$p_{1/2}$ states in row (1) are negative for three kinds of the
$ {\it YN} $ interactions, and
the results for the J\~B and NSC potentials have almost the same
order of magnitude.
The results in row (2) show that all the single-particle energies are
less attractive than those given
in row (1), and especially, those for the 0$p_{1/2}$ state are positive
(unbound) for the J\~A and NSC potentials.
This means that the Pauli-blocking effect contributes repulsively to the
single-particle energies and plays a very significant role although it
depends on the $ {\it YN} $ potential used.
In row (3) the single-particle energies become attractive in comparison
with those given in row (2) because of the restoration of
the ``diagonal" interaction.
The results in row (3) for the J\~A and NSC potentials have
almost the same order of magnitude in contrast with those given in row (1).
On the whole, we see from Table~3 that the Pauli-blocking effect is
significantly large and yields different contributions to
the single-particle energies dependently on the $ {\it YN} $ potentials.
The difference between the results shown in rows (2) and (3) is fairly large.
Therefore, we must treat the Pauli-blocking effect carefully and should
adopt the treatment of case (3) as the most accurate one.
In Fig.~4, Table~1 and Table~2,
we have displayed the results of the treatment of the Pauli-blocking effect
in case (3).

\begin{table}[b]
\caption{Contributions to the single-particle energies of $ {\it \Lambda } $
 in units of MeV. Here KE and PE stand for the
 kinetic energy and the first-order potential energy, respectively.
Rows (a), (b) and (c) are the contributions of diagrams (a),
 (b) and (c) in Fig.~2, respectively.
The quantity $ \Delta \varepsilon_{ls} $ is the spin-orbit splitting of the
0$p$ states.}
\label{table:5}
  \begin{center}
    \begin{tabular}{ccccc} \hline
        &                 &   J\~A &            &        \\
        &0$s_{1/2} $ & 0$p_{3/2}$ & 0$p_{1/2}$ &
$\Delta \varepsilon_{ls}$\\ \hline
      KE   &  10.50 &  17.50 &  17.50 &       \\
      PE   & -19.62 & -14.50 & -12.97 &  1.53 \\
     (a)   &  -1.73 &  -2.83 &  -3.03 & -0.20 \\
     (b)   &  -1.00 &  -0.43 &  -0.05 &  0.39 \\
     (c)   &  -0.33 &  -0.51 &  -0.81 & -0.31 \\
   Total   & -12.19 &  -0.78 &   0.64 &  1.41 \\
    \end{tabular}\\
    \begin{tabular}{ccccc} \hline
       &                 &   J\~B &             &        \\
        &0$s_{1/2} $ & 0$p_{3/2}$ & 0$p_{1/2}$ &
$\Delta \varepsilon_{ls}$\\ \hline
      KE   &  10.50 &  17.50 &  17.50 &       \\
      PE   & -22.44 & -16.43 & -14.81 &  1.62 \\
     (a)   &  -1.39 &  -2.19 &  -2.36 & -0.17 \\
     (b)   &  -1.31 &  -0.51 &  -0.11 &  0.40 \\
     (c)   &  -0.90 &  -1.17 &  -1.45 & -0.28 \\
   Total   & -15.54 &  -2.81 &  -1.24 &  1.57 \\
    \end{tabular}\\
    \begin{tabular}{ccccc} \hline
       &                 &    NSC       &                 &               \\
        &0$s_{1/2} $ & 0$p_{3/2}$ & 0$p_{1/2}$ &
$\Delta \varepsilon_{ls}$\\ \hline
      KE   &  10.50 &  17.50 &  17.50 &       \\
      PE   & -19.06 & -14.71 & -12.95 &  1.76 \\
     (a)   &  -1.68 &  -2.76 &  -3.04 & -0.28 \\
     (b)   &  -0.93 &  -0.52 &  -0.12 &  0.41 \\
     (c)   &  -0.52 &  -0.68 &  -0.90 & -0.22 \\ 
   Total   & -11.69 &  -1.17 &   0.49 &  1.66 \\ \hline
    \end{tabular}
  \end{center}
\end{table}

Next, the partial-wave contributions to the first-order potential
energy of $ {\it \Lambda } $ in the 0$s_{1/2}$ state are given
for the different treatments of the Pauli-blocking effect.
We see from Table~4 that the contributions of the $^{1}S_{0}$ and
$^{3}S_{1}$-$^{3}D_{1}$ channels are very different dependently on the
$ {\it YN} $ interactions employed.
In the $^{1}S_{0}$ channel the largest Pauli-blocking effect is
brought about in the result for the J\~A potential.
On the other hand, in the $^{3}S_{1}$-$^{3}D_{1}$ channel
we have an extremely large Pauli-blocking effect for the NSC potential.
This result shows that the NSC potential has the strong
$ {\it \Sigma N} $-$ {\it \Lambda N} $
coupling in the $^{3}S_{1}$-$^{3}D_{1}$ channel as discussed in Ref.~20).
The absolute values contributing to the first-order potential energy of
$ {\it \Lambda } $ in the channels with angular momenta $ l\geq 1 $
are much smaller
than those in the $^{1}S_{0}$ and $^{3}S_{1}$-$^{3}D_{1}$ channels.

In Table~5 we display the single-particle energies of $ {\it \Lambda } $
with the contributions
of the perturbation corrections given in Fig.~3, where the Pauli-blocking
effect is treated accordingly to case (3).
From this table we see that the J\~A and NSC potentials yield
almost the same values of the first-order potential energy PE and
 the perturbation corrections in spite of having different
partial-wave contributions as shown in Table~4.
The single-particle energies obtained for the J\~B potential are most
attractive among the ${\it YN} $ interactions used.
It should be noted here that the values of the KE plus PE in the 0$p$ states
for three kinds of $ {\it YN} $ interactions are positive (unbound).
Therefore, the perturbation corrections play a decisive role when we
argue that the $ {\it \Lambda } $ in the 0$p$ states is bound or not.
Correction (a) in Fig.~3 is induced by the presence of non-diagonal terms
in the one-body
hamiltonian $ h_{\sst {\it \Lambda }}=t_{\sst {\it \Lambda }}
+u_{\sst {\it \Lambda }}$.
This effect should be treated
in the Hartree-Fock (HF) calculation.
However, in the present calculation, we treat this effect perturbatively.
These non-diagonal terms work to change the single-particle wave function
of $ {\it \Lambda } $.
Therefore, this procedure is indispensable in this approach 
since we have taken a fixed value of $ {\it \hbar \Omega } $ commonly for
$ {\it \Lambda } $ and $ {\it N} $.
This correction~(a) contributes most attractively to
the single-particle energies among all the corrections in Fig.~3.
The magnitude of every correction is considerably small in comparison with the
first-order potential energies for all the $ {\it YN} $ interactions used.
All the perturbation corrections contribute, however, attractively to
the single-particle
energies of $ {\it \Lambda } $, and do not change the mutual spacings of
the single-particle states significantly. 
This trend is very different from the situation observed in 
ordinary nuclei such as $^{16}$O and $^{40}$Ca,
in which the correction terms
 contribute to about 40$\%$ of the calculated spin-orbit splitting of
 the 0$p$ or 0$d$ states \cite{rf:SO,rf:KSO}.
The role of many-body correlations, such as the core polarization, does not
seem to be important in the spin-orbit splitting of $ {\it \Lambda } $.

\section{Concluding remarks}
\setcounter{equation}{0}

We have presented a formulation of the UMOA
to apply it to a structure calculation of hypernuclei.
We have introduced a unitary transformation exp($S$),
with the correlation operator $S$,
that describes two-body correlation of hyperon and
nucleon.
An equation has been given for determining the
correlation operator $S$ as the equation of
decoupling so that the unitarily transformed interaction does not
have non-zero matrix elements between the model space $P$ of
low-lying $ {\it \Lambda N} $ states and its complement $Q$.
The effective interaction has been given
by the projected interaction onto the $P$ space.
The effective interaction thus determined has properties of being hermitian,
energy-independent and decoupled between the $P$ and $Q$ spaces.
In comparison with the usual $G$-matrix approach, the use of such an
effective interaction would have some advantages in describing
many-body systems.

The UMOA has been applied to the calculation of the structure
of $_{\it \Lambda }^{17}$O. The $ {\it \Lambda N} $ effective
interaction has been
calculated by employing three ${\it YN}$ potentials, namely, the
J\~A, J\~B and NSC potentials.
With the $ {\it \Lambda N} $ effective interaction,
we have calculated the single-particle energies
of $ {\it \Lambda } $ in the 0$s_{1/2}$, 0$p_{3/2}$ and 0$p_{1/2}$ states.

We have seen that how to treat the Pauli-blocking effect is
very important, especially when we solve the problem of the
$ {\it \Sigma N} $-$ {\it \Lambda N} $ coupling.
In the usual treatment one excludes all the interactions acting in a space
of  $ {\it {\Lambda}N} $ and $ {\it {\Sigma}N} $ states in which a nucleon is in occupied states.
However, this treatment of the Pauli-blocking effect would lead to
overmuch counting.
We have introduced a new treatment in which, while we exclude the interactions
acting among forbidden $ {\it YN} $ states as in the usual treatment,
we restore the interactions that are diagonal in occupied nucleon states.
We have observed that this treatment of the Pauli-blocking effect
has given rise to a significant effect as compared with the usual treatment.

The present results have shown that $ {\it \Lambda } $ in the 0$p_{1/2}$
state is unbound for the J\~A and NSC potentials and,
on the other hand, bound for the J\~B potential.
The spin-orbit splitting of the single-particle levels
of ${\it \Lambda}$ in the 0$p$ states has been given by 1.41,
1.57 and 1.66 for the J\~A, J\~B and NSC
potentials, respectively.
These values are
very large compared with the results in Ref.~\cite{rf:VID98}. It will be
of high interest for the single-particle nature  of $ {\it \Lambda } $
in $_{\it \Lambda }^{17}$O to be established experimentally.

In the application of the UMOA some problems have still
remained. The unitarily transformed hamiltonian \cite{rf:SO} contains
originally three-or-more-body interactions. The effects of these
many-body interactions have been evaluated in the previous works
of calculating nuclear properties, and we have concluded that
the contributions of these many-body terms are much smaller
than those of two-body interactions but sizable in some cases.
We therefore should estimate the effects of these many-body
interactions also in hypernuclei.   

Another refinement to be made is that, in the present 
calculation, the single-particle energies of $ {\it \Sigma } $
in intermediate states are neglected.
However, the single-particle spectrum of $ {\it \Sigma } $ has a
possibility of giving rise to some contributions to the
$ {\it \Lambda N} $ effective interaction. Although the properties
of the single-particle energies of $ {\it \Sigma } $ have not always
been made clear,
we should investigate how much the spectrum of $ {\it \Sigma } $ affects
in the determination of the $ {\it \Lambda N} $ effective interaction.
This problem is also a remaining task in the UMOA.

\section*{Acknowledgements}
The authors would like to thank Y.~Yamamoto and K.~Miyagawa
for useful discussions and their help for using the Nijmegen and
J\"ulich ${\it YN} $ potential codes.
One of the authors (S.~F.) would like to express his thank to
K.~Takada, M.~Kamimura and Y.~R.~Shimizu for
their continuous encouragement and instructive discussions.

This work is supported in part by a Grant-in-Aid for Scientific Research from
 the Ministry of Education, Science, Sports and Culture, Japan (No. 09225204).

\end{document}